\documentclass{article}

\usepackage{empheq}
\usepackage{amssymb}
\usepackage{amsthm}
\usepackage{fancyhdr}
\usepackage{authblk}
\usepackage[title]{appendix}
\usepackage{array}
\usepackage[tableposition=top]{caption}
\usepackage{dsfont}
\usepackage{graphicx}
\usepackage{mathtools}
\usepackage{multicol}
\usepackage{multirow}
\usepackage[square]{natbib}
\usepackage{pgfplots}
\usepackage{tikz}
\usepackage{xcolor}
\usepackage{enumitem}
\setlist[itemize]{noitemsep}
\setlist[enumerate]{noitemsep}
\usepackage{arxiv}
\usepackage{booktabs}       % professional-quality tables
\usepackage{nicefrac} 
\usepackage{tabularx}

\usepackage{enumitem}
\usepackage{algorithm}
\usepackage{algpseudocode}
\usepackage{xpatch}
\usepackage{subfig}

\pgfplotsset{compat=1.10}

\usetikzlibrary{pgfplots.statistics}
\pgfplotsset{compat=1.12}
\usepackage{graphicx}
\usetikzlibrary{}

\usepgfplotslibrary{statistics}

\usepackage[OT1]{fontenc}

\usepackage{url}
\usepackage{hyperref}

\usepackage[capitalise]{cleveref}
\usepackage{oubraces}
\usepackage{amsmath}
\usepackage{arxiv}

\pgfplotsset{compat=1.4}

\pgfplotsset{
    jitter/.style={
        x filter/.code={\pgfmathparse{\pgfmathresult+(rnd-.5)*#1}}
    },
    jitter/.default=0.5
}

\usetikzlibrary{hobby, graphs, decorations.markings, arrows, shapes, positioning, mindmap, backgrounds, calc, matrix, patterns, bayesnet, patterns.meta, shapes.geometric, arrows.meta, fit, tikzmark, arrows.meta}

\usepgfplotslibrary{statistics, fillbetween, external}

\makeatletter
\xpatchcmd{\algorithmic}{\itemsep\z@}{\itemsep=0.8ex plus2pt}{}{}
\makeatother

\makeatletter
\renewcommand{\fnum@algorithm}{\fname@algorithm}
\makeatother
\algrenewcommand\algorithmicrequire{\textbf{Input:}}
\algrenewcommand\algorithmicensure{\textbf{Output:}}
\algnewcommand{\LeftComment}[1]{\Statex \(\triangleright\) #1}

\usepackage[table]{xcolor}

\usepackage{pgfplots}  
\usepackage{pgfplotstable}  
\pgfplotsset{compat=1.17}  

\usetikzlibrary{hobby, graphs, decorations.markings, arrows, shapes, positioning, mindmap, backgrounds, calc, matrix, patterns, bayesnet, patterns.meta}

\usepgfplotslibrary{statistics, fillbetween, external}

\newtheorem{theorem}{Theorem}

\newtheorem{definition}{Definition}

\newtheorem*{proof*}{Proof}

\newtheorem{assumption}{Assumption}

\newtheorem{remark}{Remark}

\newcommand{\ind}{\mathds 1}

\newcommand{\indep}{\protect{\perp \!\!\! \perp}}

\newcommand{\nA}{{N_{ \mathcal A }}}

\newcommand{\nB}{{N_{ \mathcal B }}}

\newcommand{\KR}[1]{\textcolor{black}{#1}} % [Kayané:\ #1]

\begin{document}

\large

\title{ Causal Record Linkage }

\renewcommand\Authsep{, }
\renewcommand\Authands{, and }

\author{Kayan\'{e} Robach\thanks{Corresponding Author: \texttt{k.c.robach@amsterdamumc.nl}}}

\author{Michel H.\ Hof}

\author{Mark A.\ van de Wiel}

\author{St\'{e}phanie L.\ van der Pas}

\affil{Department of Epidemiology and Data Science, \protect\\ Amsterdam UMC location Vrije Universiteit Amsterdam, \protect\\ De Boelelaan 1117, 1081 HV Amsterdam, The Netherlands}
\affil{Amsterdam Public Health, Methodology, The Netherlands}
\affil{Department of Mathematics, Vrije Universiteit Amsterdam, \protect\\ De Boelelaan 1111, 1081 HV Amsterdam, The Netherlands} 

\maketitle

\abstract{ 
Combining observations from multiple data sources creates a valuable resource for research. In causal inference, record linkage methods grant researchers access to the covariates needed for unbiased estimation and offer the means to explore (long-term) outcomes. This paper examines how record linkage conflicts with the assumptions required for identifying causal effects. Our investigation reveals that linkage errors primarily threaten consistency in the causal framework, leading to attenuation bias in the inference due to discrepancies in the linked data. In attempting to address this bias by being more stringent on the linkage, \KR{sampling bias inadvertently emerges from a covariate shift between the baseline and the linked samples,} as profiles with more distinctive information are more likely to be recovered. Despite these challenges, we identify settings in which record linkage and causal inference can be soundly combined, and show how the effect estimated on linked data can be generalised, with linkage decisions informed accordingly. We further identify when existing and novel solutions support valid causal inference on linked data, and when inference should instead be treated with caution or even abandoned. We illustrate the challenges raised and the potential solutions using real data from a study of women's health and a simulation study. 
}


\section{Introduction} \label{sec introduction}

The extent to which reliable causal inference can be achieved on linked data is ambiguous, as linkage processes can compromise the validity of common causal assumptions. In causal inference, baseline information and exposure may come from a prior study while outcomes are collected later. A lack of unique identifiers is a major obstacle towards combining individual records across heterogeneous data sets, and record linkage methods address this by connecting information on the same individuals using partially identifying variables (e.g.\ birth year, postal code, ethnicity), thereby offering access to broader sets of variables across wider populations and extended time periods. When integrating data through record linkage, it is important to assess the implications on the causal framework and to establish guidelines for selecting records to ensure reliable causal estimates.

In this paper, we consider the causal record linkage task as integrating two data sources to construct a study population defined by their intersection, in order to perform causal inference. Record linkage relies on pre-treatment linking variables $\boldsymbol{Z}$, observed in both sources, to recover complete profiles by linking pre-treatment covariates $\boldsymbol{X}$ and binary treatment $T$ from one source with outcome $Y$ from another, inferred to correspond to the same entities. Statistical inference is then conducted on the conditional distribution of $Y|T = t, \boldsymbol{X}$ for $t \in \{0,1\}$, for the imperfect selected set of complete profiles built through record linkage, either jointly \citep{shan_joint_CRL, guha_bay_causal_RL_2022, guha_reiter_cov_joint} or via a two-stage approach \citep{wortman_RL_causality_2018, wortman_rl_causal, luo2023multiplyrobustestimationcausal}. 

There are two main paradigms in record linkage: the Fellegi-Sunter model and the hierarchical entity resolution framework \citep{FS_RL_1969, tancredi_bay_RL_2011, marchant_bayER_2023} (we illustrate both along this article). All methods essentially output a bimodal distribution of linkage scores, expressing the amount of evidence that two observations belong to the same individual, taking values in $[0, 1]$ \citep{jaro_proba_linkage_1995, twins_rl_appl2008, review_methods_2014, rl_Appl_2018, robachflexRL}. The modes distinguish obviously distinct profiles (low scores) from similar ones (high scores). A cutoff $\lambda \in [0.5, 1)$ is applied to these scores as a decision rule, classifying pairs to yield a valid set of linked records \citep{tancredi_bay_RL_2011, sadinle_bay_bipartite_RL_2017}. The choice of this cutoff depends on the goal of the analysis, ideally minimising the false discoveries or tailored to the inference objective \citep{wortman_RL_causality_2018}.

Linking variables $\boldsymbol{Z}$ are often demographics \citep{Hershberg01061976, WINKLER2009351, review_methods_2014, demogr_rl} and may therefore be causally relevant (e.g.\ sex, age, socioeconomic status, geographic region). For that reason, linking variables may be involved in the inference, and thus in the treatment effect heterogeneity. We deviate from the literature \citep{wortman_RL_causality_2018, guha_bay_causal_RL_2022} that ignored the overlap between $\boldsymbol{Z}$ and $\boldsymbol{X}$, and hence did not investigate the issues induced by record linkage on causal effect estimation \citep{reflexRL_2020, linkage_bias_datasynthesis24}. Our framework extends the literature on the implications of record linkage on statistical inference \citep{Lahiri_Larsen_linked_2005, Tancredi_Liseo_linked_2015, reflexRL_2020, Wang_Slawski_linked_2022} to causal inference. The conditions under which causal inference can be credibly performed on linked data in such a realistic setting have not been examined, yet as we shall demonstrate in this paper, covariate dependent linkage errors interfere with the causal framework and can introduce noteworthy biases in the causal effect estimation.

Causal inference typically rests on three core assumptions \citep{pearl_2009, imbens_causal_2015, whatif, li_bayci_2023}: consistency between observed and potential outcomes, positive probability of exposure for all profiles and conditional exchangeability between treatment groups. \KR{We explain how record linkage errors that depend on the covariates affect these assumptions. Falsely linked records corrupt consistency, threatening the internal validity of the linked sample. Besides, record linkage preferentially selects distinctive profiles, which shifts the covariate distribution away from the original population, limiting the external validity of causal conclusions. We quantify the resulting biases through a decomposition into attenuation and sampling bias, and establish conditions under which the effect in the original population is identifiable and estimable from the linked sample.}

Motivated by the peculiar linkage sample selection mechanism, the main methodological contributions of this paper are to formalise the task of causal inference on linked data in \Cref{sec population}, to discuss the qualitative implications on effect identifiability in \Cref{sec identifiability} and to address the transportability of findings despite the selection. We quantify the linkage induced biases in \Cref{sec bias decompo} and propose solutions for estimating treatment effects with linked data in \Cref{sec estimation}. In \Cref{sec linkage selection} we provide diagnostic tools for causal record linkage reliability. We illustrate this work with real data from an American study of women's health through \Cref{sec application}, and with a simulation study in \Cref{sec simulations}. The results presented in this paper are independent of the record linkage method used.

\section{Causal Record Linkage and Sample Selection Bias} \label{sec population}

The growing availability of observational data has driven advances in causal inference, yet data access remains restricted \citep{data_causal}. Record linkage combines pseudonymised data across sources, enabling the study of novel causal questions. In this section we present the record linkage setting alongside the causal inference problem at stake, before examining the impact of the linkage decisions on the causal roadmap.

Record linkage considers two overlapping samples from the same \emph{source population} $\mathcal{P}$, aiming to construct a study population defined by their intersection: the set of \emph{(true) links} i.e.\ pairs of records belonging to the same individuals. The \emph{baseline data} set $\mathcal{A}$ gathers baseline characteristics $\boldsymbol{C}$ and binary treatment $T$, while the \emph{follow-up data} set $\mathcal{B}$ gathers outcome $Y$. \begin{assumption} \label{ass linkage random} $\mathcal{A}$ and $\mathcal{B}$ are overlapping random samples from $\mathcal{P}$. \end{assumption} Implicitly, $\mathcal{P}$ must not be too large relative to $\mathcal{A}$ and $\mathcal{B}$, ensuring a sufficiently likely overlap between samples. The intersection of data sources can then be regarded as a random sample from the collected data \citep{tancredi_bay_RL_2011, marchant_bayER_2023, luo2023multiplyrobustestimationcausal, robachflexRL}, and any unit in $\mathcal{A}$ is expected to be followed up in $\mathcal{B}$. Without unique identifiers, observations are linked based on pre-treatment variables $\boldsymbol{Z}$, common to both sources and possibly overlapping with $\boldsymbol{C}$. \begin{assumption} \label{key ass hold} The key assumptions for causal identification hold in the source population: consistency, positivity, conditional exchangeability. \citep{imbens_causal_2015, whatif} \end{assumption} We expand on these in \Cref{sec identifiability}. Consequently, all units in $\mathcal{A}$ are eligible for treatment; we denote the propensity score as a function of the covariates: $\pi(\boldsymbol{X}) = \mathbb P (T=1|\boldsymbol{X})$. Throughout the paper, index $i$ refers to information from $\mathcal{A}$, $j$ to information from $\mathcal{B}$, and $N$ indexed by a set denotes its size. The observed data are given by: $$\mathcal{A}=\big\{(T_i,\boldsymbol{Z}_i,\boldsymbol{C}_i) \bigm| i=1,\dots,\nA\big\},\ \ \ \mathcal{B}=\big\{(\boldsymbol{Z}_j,Y_j) \bigm| j=1,\dots,\nB \big\}.$$

\subsection{Sample selection} \label{sec sample selection}

\begin{figure}[!ht]
    \centering
    \begin{tikzpicture}
    \pgfplotsset{
    width=0.5\textwidth,
    height=0.23\textwidth}
    \begin{axis}[
        ymin=0, ymax=8,
        ytick={0,2,4,6,8},
        yticklabels={,,,,},
        ylabel={\textit{BRL}},
        xtick style={opacity=0}, 
        xtick={0, 0.5, 0.625, 0.675, 0.735, 1},
        xticklabels={0, 0.5, $\overset{}{\leftarrow}$, $\lambda$, $\overset{}{\longrightarrow}$, 1}, legend style={fill=none, draw=none}
        ]
    % BRL link score given to (true) links
    \addplot+[ybar interval, mark=no, draw=black, fill=black!70, area legend
    ] plot coordinates { (0.00,2.66) (0.05,0.00) (0.10,0.03) (0.15,0.39) (0.20,0.98) (0.25,1.13) (0.30,1.30) (0.35,1.19) (0.40,1.34) (0.45,1.65) (0.50,1.50) (0.55,1.29) (0.60,1.08) (0.65,1.01) (0.70,1.16) (0.75,1.46) (0.80,1.83) (0.85,1.97) (0.90,2.23) (0.95,2.19) (1.00,2.19) };
    % BRL link score given to all records
    \addplot+[ybar interval, mark=no, draw=black!70, fill=black!50, area legend, opacity=0.6
    ] plot coordinates { (0.00,6.41) (0.05,0.00) (0.10,0.50) (0.15,1.29) (0.20,1.60) (0.25,1.73) (0.30,1.78) (0.35,1.70) (0.40,1.70) (0.45,1.98) (0.50,1.79) (0.55,1.51) (0.60,1.25) (0.65,1.11) (0.70,1.28) (0.75,1.58) (0.80,1.90) (0.85,2.05) (0.90,2.26) (0.95,2.20) (1.00,2.20) };
    \legend{true links, other records pairs}
    \end{axis}
    \end{tikzpicture} \hspace{10pt}
    \begin{tikzpicture}
    \pgfplotsset{
    width=0.5\textwidth,
    height=0.23\textwidth}
    \begin{axis}[
        ymin=0, ymax=8,
        ytick={0,2,4,6,8},
        yticklabels={,,,,},
        ylabel={\textit{FlexRL}},
        xtick style={opacity=0},
        xtick={0, 0.5, 0.625, 0.675, 0.735, 1},
        xticklabels={0, 0.5, $\overset{}{\leftarrow}$, $\lambda$, $\overset{}{\longrightarrow}$, 1}, legend style={fill=none, draw=none}
        ]
    % FlexRL link score given to (true) links
    \addplot+[ybar interval, mark=no, draw=black, fill=black!70, area legend
    ] plot coordinates { (0.00,2.50) (0.05,1.73) (0.10,1.64) (0.15,1.55) (0.20,1.40) (0.25,1.35) (0.30,1.32) (0.35,1.44) (0.40,1.50) (0.45,1.52) (0.50,1.34) (0.55,1.16) (0.60,0.98) (0.65,0.87) (0.70,0.82) (0.75,1.32) (0.80,1.74) (0.85,2.04) (0.90,2.07) (0.95,2.33) (1.00,2.33) };
    % FlexRL link score given to all records
    \addplot+[ybar interval, mark=no, draw=black!70, fill=black!50, area legend, opacity=0.6
    ] plot coordinates { (0.00,6.41) (0.05,2.63) (0.10,2.31) (0.15,2.10) (0.20,1.87) (0.25,1.74) (0.30,1.67) (0.35,1.73) (0.40,1.80) (0.45,1.82) (0.50,1.62) (0.55,1.41) (0.60,1.15) (0.65,1.02) (0.70,0.96) (0.75,1.43) (0.80,1.86) (0.85,2.12) (0.90,2.12) (0.95,2.35) (1.00,2.35) };
    \legend{true links, other records pairs}
    \end{axis}
    \end{tikzpicture}
    % \begin{tikzpicture}
    % \pgfplotsset{
    % width=0.38\textwidth,
    % height=0.25\textwidth}
    % \begin{axis}[
    %     ymin=0, ymax=3,
    %     ytick={0,2},
    %     yticklabels={,},
    %     xtick={0.5,0.58,0.625,0.68,1},
    %     xticklabels={0.5,$\overset{}{\leftarrow}$,$\lambda$,$\overset{}{\longrightarrow}$,1},
    %     ]
    % \addplot+[ybar interval, mark=no, draw=black
    % ] plot coordinates { (0.50,1.08) (0.55,0.70) (0.60,0.48) (0.65,0.60) (0.70,0.85) (0.75,0.95) (0.80,0.78) (0.85,0.60) (0.90,0.30) (0.95,0) (1,0)};
    % \addplot+[ybar interval, mark=no, draw=gray!70, fill=gray!50
    % ] plot coordinates { (0.50,0.40) (0.55,0.20) (0.60,0.00) (0.65,0.50) (0.70,0.75) (0.75,0.85) (0.80,0.78) (0.85,0.60) (0.90,0.30) (0.95,0) (1.00,0) };
    % \addplot[
    % dashed,
    % black
    % ] coordinates {(0.625,0) (0.625,3)};
    % \end{axis}
    % \end{tikzpicture}
    \caption{Stacked linkage scores for all records pairs of the SWAN data, obtained from two different methods, \textit{BRL} \citep{sadinle_bay_bipartite_RL_2017} (left) and \textit{FlexRL} \citep{robachflexRL} (right). Scores (in pale) express the amount of evidence that observations belong to the same entities. To limit contamination from falsely linked data, we select pairs with scores beyond $\lambda \in [0.5,1)$, hoping to mostly detect true links (with scores in dark). Lower $\lambda$ yields more falsely linked and higher $\lambda$ more missed links.}
    \label{fig:LinkScore}
\end{figure}
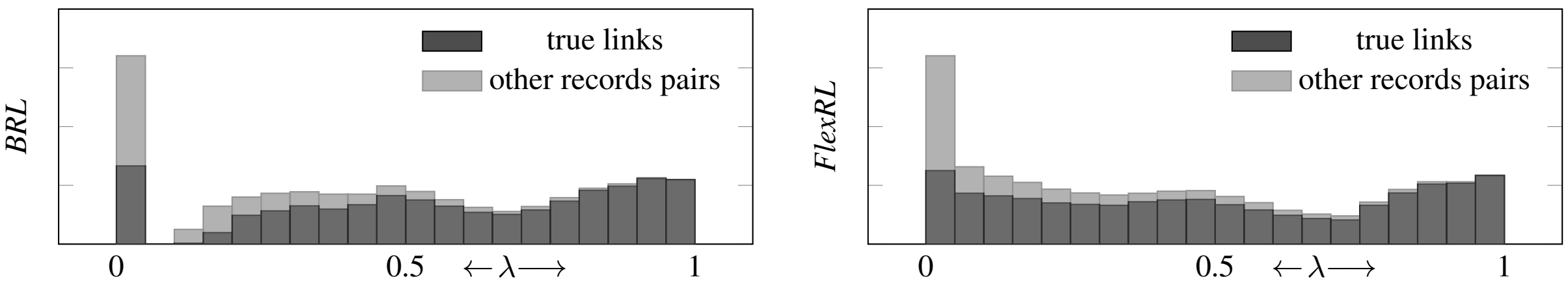

Notwithstanding the opportunity to address novel causal questions with linked data, record linkage restricts the sample to a selected population whose covariate distribution likely differs from that of the source population \citep{sample_selec_79, reflexRL_2020, RLoverlooked}. This selection stems from imperfect identifying information, resulting in missed links between entities present in both sources and falsely linked records from distinct entities. Being more stringent by raising the cutoff $\lambda$ reduces the latter but increases the former. In practice, the true links may be difficult to distinguish, as indicated by the superposition of linkage scores in \Cref{fig:LinkScore}, obtained with two state-of-the-art algorithms on the application of \Cref{sec application}. Links recovered with high confidence tend to involve individuals with particularly distinctive identifying information, yielding a sample biased towards atypical profiles: record linkage acts as a selection mechanism.

Related works have considered record linkage for causal inference when identifying information is independent of causally relevant variables, assuming that linkage does not interfere with the causal framework, or restricting inference to homogeneous treatment effects \citep{wortman_RL_causality_2018, guha_bay_causal_RL_2022}. Since the linking variables may overlap with those driving treatment assignment and outcome, we allow them to be causally relevant and consider $\boldsymbol{X}$ as the union of $\boldsymbol{Z}$ and $\boldsymbol{C}$, extending the literature towards effect heterogeneity and exploring how linkage errors interact with the causal framework.

More broadly, this connects to the data fusion literature, where transportability across domains may introduce biases \citep{Robins_1986_g, stuart_ipw, cole_g, bareinboimpearl2016, dahabreh_transport, colnetetal_2024}. The structural causal framework \citep{SEM_1943, SCM_1995} formalises such population differences via a selection variable $S$, reflecting implicit or explicit conditioning on study enrolment, affecting the distribution of variables of interest \citep{bareinboim12GAdmissibility, BareinboimTianPearl2014, pearlbareinboim2014, bareinboimpearl2016, colnetetal_2024}. We borrow this machinery and introduce a graphical causal model on \Cref{sec model} to understand how common causal assumptions interact with record linkage in \Cref{sec identifiability}.

\subsection{Graphical causal model} \label{sec model}

\begin{figure}[!ht]
    \centering
    \subfloat[\small Causal graph with Selection (source population, $\mathcal{P}$) \label{subfig:Event1}]{
    \begin{tikzpicture}[
      every node/.style={align=center},
      assumption/.style={draw, circle},
      check/.style={draw, rounded corners},
      result/.style={draw},
      arrow/.style={->, thick}
    ]
    \node[shape=circle, draw=black, minimum size=1cm] (x) at (-0.25,1) {$\boldsymbol{X}$};
    \node[shape=circle, draw=black, minimum size=1cm] (t) at (-3,-3) {$T$};
    \node[shape=rectangle, draw=black, minimum size=1cm] (s) at (-0.25,-1.82) {$S$};
    \node[shape=circle, draw=black, minimum size=1cm] (y) at (2.5,-3) {$Y(t)$};
    \node[shape=circle, minimum size=1cm] (transparent) at (3,-3.35) {};
    
    \node[draw,shape=circle,minimum size=0.5cm] (tt) at (-2.18,-3) {$t$};

    % Directed edge
    \path [-stealth] (x) edge (s);
    \path [-stealth] (x) edge (t);
    \path [-stealth] (x) edge (y);
    \path [-stealth] (tt) edge (y);
    \end{tikzpicture} } \hspace{20pt} \subfloat[\small Causal graph for linked pairs (selection biased sample, $\mathcal{S}$) \label{subfig:Event2}]{ \begin{tikzpicture}[
      every node/.style={align=center},
      assumption/.style={draw, circle},
      check/.style={draw, rounded corners},
      result/.style={draw},
      arrow/.style={->, thick}
    ]
    \node[shape=circle, draw=black, minimum size=1cm] (xi) at (-0.25,1) {$\boldsymbol{X}_i$};
    \node[shape=circle, draw=black, minimum size=1cm] (ti) at (-3,-3) {$T_i$};
    \node[shape=rectangle, draw=black, minimum size=1cm] (s) at (0.4375,-2.2) {$S_{i,j}$};
    \node[shape=circle, draw=black, minimum size=1cm] (yi) at (2.5,-3) {$Y_i(t)$};
    
    \node[draw,shape=circle,minimum size=0.5cm] (tti) at (-2.18,-3) {$t$};

    \node[shape=circle, draw=black, minimum size=1cm] (xj) at (1.125,0.3) {$\boldsymbol{X}_j$};
    \node[shape=circle, draw=black, minimum size=1cm] (tj) at (-1.625,-3.7) {$T_j$};
    \node[shape=circle, draw=black, minimum size=1cm] (yj) at (3.875,-3.7) {$Y_j(t)$};
    
    \node[draw,shape=circle,minimum size=0.5cm] (ttj) at (-0.805,-3.7) {$t$};

    % Directed edge
    \path [-stealth] (xi) edge (s);
    \path [-stealth] (xi) edge (ti);
    \path [-stealth] (xi) edge (yi);
    \path [-stealth] (tti) edge (yi);
    \path [-stealth] (xj) edge (s);
    \path [-stealth] (xj) edge (tj);
    \path [-stealth] (xj) edge (yj);
    \path [-stealth] (ttj) edge (yj);
    \end{tikzpicture} }
    
    \caption{Selection diagrams depicting differences between source population contained in the data and linked population obtained with record linkage (indicated by $S$). The selection process is made on the linking variables $\boldsymbol{Z}$ included in $\boldsymbol{X}$, hence $S$ descends from $\boldsymbol{X}$.}
    \label{fig:SelectionDiagram}
\end{figure}
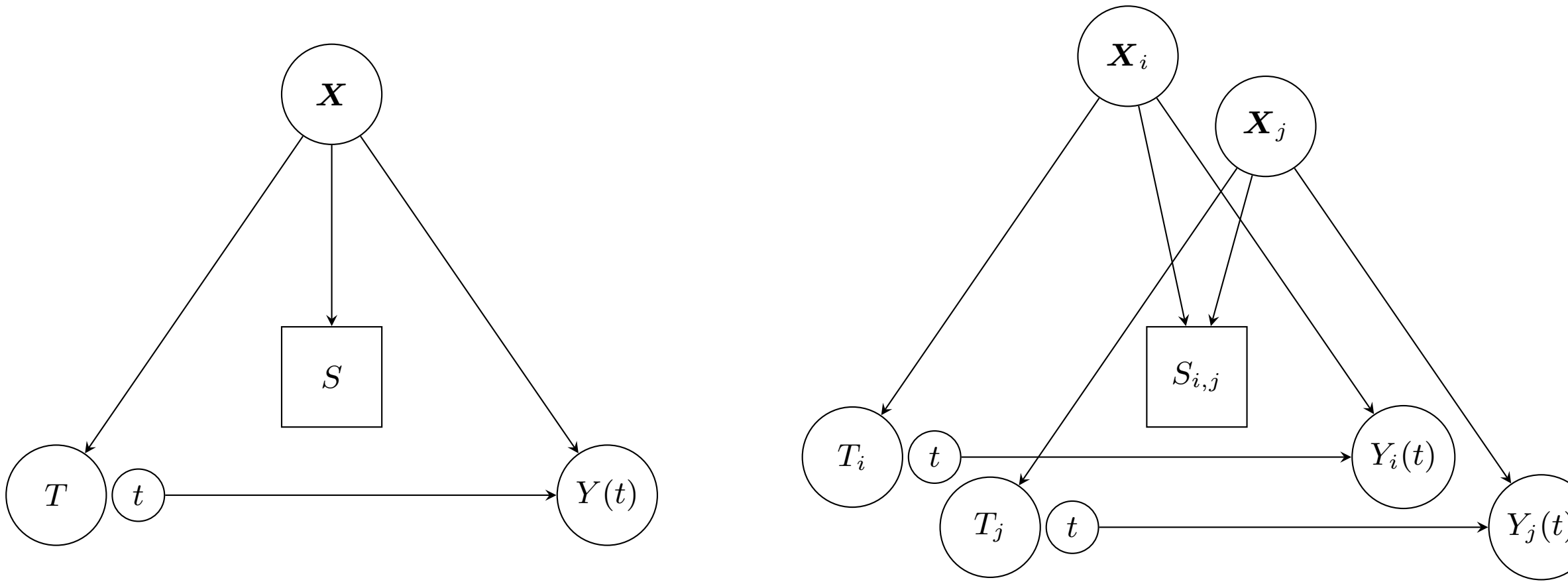

Let $S$ denote the linkage estimator, with $S_{i,j}=1$ meaning that pair $(i,j)$ has a linkage score above $\lambda$ and is declared `linked'. Inclusion in the analysis depends on linkage, which occurs differentially with respect to $\boldsymbol{X}$, making $S$ a descendant of $\boldsymbol{X}$ on \Cref{fig:SelectionDiagram}. In other words, linkage errors depend on covariates (atypical profiles are less likely to be missed or falsely linked). Yet, because linkage is not informative about the variables of interest in the analysis, $S$ has no descendants.

We represent factual and counterfactual (observed and potential) outcome distributions using single world intervention graphs (SWIGs) \citep{richardson2013swigs}. For unit $i \in \mathcal{A}$, the factual outcome is $Y_i = f_Y(\boldsymbol{X}_i,T_i,\varepsilon_{{Y}_i})$ (analogously for $j \in \mathcal{B}$), and for every unit in $\mathcal{P}$, the counterfactual outcome is $Y(\boldsymbol{x}, t)=f_Y(\boldsymbol{x},t,\varepsilon_{Y})$ for any $\boldsymbol{x}$, $t \in \{0,1\}$, sometimes denoted $Y(t)$, although substituting $Y_j$ for $Y_i(t)$ is warranted only when records $i$ and $j$ coincide on both treatment and covariates. 

\Cref{subfig:Event1} depicts the causal structure for the effect of $T$ on $Y$ confounded by $\boldsymbol{X}$, valid for i.i.d.\ units in $\mathcal{P}$ and for true links (subscripts omitted). Since linkage and causal mechanisms are intertwined, causal pathways must be considered at the level of linked pairs $\mathcal{S}$; \Cref{subfig:Event2} unravels this representation to make relationships explicit across both data sources, where units become pairs of linked entities forming an i.i.d.\ sample of selection biased data from $\mathcal{P}$.

Independencies can be tested via d-separation (directional flow of information): a path is blocked if it contains a collider that is not conditioned on (nor on any of its descendants), or a non-collider that is conditioned on; two variables are d-separated (by a set), and hence (conditionally) independent, if all paths between them are blocked \citep{SCM_1995, mit_causal_2001, pearl_2009, ogburn_14, mooij_lecture_casuality}.

On \Cref{subfig:Event2}, the independence between pairs breaks once we condition on $S_{i,j}$. Indeed, if records $i$ and $j$ refer to different individuals, they are independent observations: $\boldsymbol{X}_i \indep Y_j(t)$ for $t \in \{0,1\}$, since the only path between them, $\{\boldsymbol{X}_i \rightarrow S_{i,j} \leftarrow \boldsymbol{X}_j \rightarrow Y_j(t)\}$, is blocked by the unconditioned collider $S_{i,j}$. Once linked however, spurious dependencies emerge: $\boldsymbol{X}_i \not\!\!\!\indep Y_j(t) | S_{i,j}=1$. By involving covariates in the linkage process, the latter becomes prone to confirmation bias, as estimation indirectly favours pairs whose covariates align with the relationship between $\boldsymbol{X}$ and $Y$ under study. Understanding these implications is important to appropriately conduct inference on the potentially imperfect selected set of linked data: $$ \mathcal{S} = \big\{(\boldsymbol{X}_i,T_i,Y_j) \bigm| S_{i,j}=1, i=1,\dots,\nA, j=1,\dots,\nB \big\}. $$

\subsection{Estimand} \label{sec estimand}

An estimand defines the intended target of inference, but feasible estimands are constrained by the study design, estimation procedure, data and, critically here, by the selection mechanism induced by record linkage. We characterise the target of record linkage, and its locus in our framework, to establish what the causal estimand can feasibly aim at, given the sample selection and causal model developed above.

Let $\Delta$ denote the true link indicator, target of record linkage, with $\Delta_{i,j}=1$ if records $i$ and $j$ correspond to the same individual. Among $N_{S} = \sum_{i,j} S_{i,j}$ linked entities selected for causal inference, true positives $\text{TP} = N_{S\Delta} = \sum_{i,j} S_{i,j} \Delta_{i,j}$ are formed of records referring to the same individual, satisfying $T_i=T_j$ and $\boldsymbol{X}_i=\boldsymbol{X}_j$, while false positives $\text{FP} = N_{S\nabla} = \sum_{i,j} S_{i,j} (1-\Delta_{i,j})$ arise from records with similar profiles pertaining to distinct individuals. The remaining true links are missed as false negatives, $\text{FN} = \sum_{i,j} (1-S_{i,j}) \Delta_{i,j}$. These yield linkage quality metrics: False Discovery Proportion (FDP) $\frac{FP}{FP+TP} = \frac{N_{S\nabla}}{N_{S}}$ and sensitivity $\frac{TP}{FN+TP}$. 

Under \Cref{ass linkage random}, the distributions of variables of interest are identical across true links ($\Delta_{i,j}=1$) and non-links ($\Delta_{i,j}=0$), which translates into $(\boldsymbol{X}, T, Y) \indep \Delta$. More generally, the latent $\Delta$ plays no role in generating dependence among $(\boldsymbol{X}, T, Y, S)$, it is therefore omitted from our causal graphical model. \begin{assumption} \label{ass delta separated} $\Delta$ is disconnected from the system: $\Delta \indep V | W$ for any $V, W \subseteq \{\boldsymbol{X}, T, Y, S\}$. \end{assumption} Under Assumptions \ref{ass linkage random} and \ref{ass delta separated}, TP and FP share the same distribution of variables of interest, $(\boldsymbol{X}, T, Y) \indep \Delta|S$. These independencies imply $\boldsymbol{X} \indep \Delta$ and $\boldsymbol{X} \indep \Delta |S$, yet $\boldsymbol{X} \not\!\!\!\indep S$, with implications for the causal estimand.

\begin{remark} \label{ass same cov selec}
    Under \Cref{ass linkage random}, $\boldsymbol{X}_i$ and $\boldsymbol{X}_j$ share the same marginal distribution and, registration processes of $\mathcal{A}$ and $\mathcal{B}$ are comparable (same patterns of mistakes and missing values), implying for any $\boldsymbol{x}$ that $\mathbb P (S_{i,j}=1|\boldsymbol{X}_i=\boldsymbol{x})=\mathbb P (S_{i,j}=1|\boldsymbol{X}_j=\boldsymbol{x})$, meaning in this context that $\boldsymbol{X}_i|S_{i,j} \sim \boldsymbol{X}_j|S_{i,j}$.
\end{remark}

Linkage errors (FP and FN) bias inference \citep{Lahiri_Larsen_linked_2005, Tancredi_Liseo_linked_2015, reflexRL_2020, Wang_Slawski_linked_2022} and threaten causal validity as \Cref{sec identifiability} will show. \KR{Falsely linked records undermine the internal validity of the linked data which precludes estimands targeting a specific sample (e.g.\ individual treatment effect or average treatment effect among true links) and confines inference to estimands targeting the source population. Yet this population is only partially observed through the linked data, which raises concerns about their external validity.}

Throughout this article, the potential outcome mean and conditional mean under intervention $T=t$ are denoted $\mu_t = \mathbb E [Y(t)]$ and $\mu_t(\boldsymbol{X}) = \mathbb E [Y(t)|\boldsymbol{X}]$ respectively. In contrast, the observed outcome under treatment $t \in \{0,1\}$ is denoted $Y|T=t$. We focus on the Average Treatment Effect (ATE) in the baseline population: \begin{align} \label{estimand_tau_A}
    \tau_{\mathcal{A}} = \mathbb E_{\mathcal{A}} [Y(1)-Y(0)] = \mathbb E_{\mathcal{A}} [\, \mathbb E [Y(1)-Y(0) | \boldsymbol{X}] \,] = {\mu_1}_{\mathcal{A}} - {\mu_0}_{\mathcal{A}} = \mathbb E [ {\mu_1}_{\mathcal{A}}(\boldsymbol{X}) - {\mu_0}_{\mathcal{A}}(\boldsymbol{X}) ].
\end{align} The emphasis on the covariate domain in the expectations echoes some considerations regarding potential outcomes to be expanded in \Cref{sec consistency}. \KR{We will now relate the target $\tau_{\mathcal{A}}$ to the observed data through the sample ATE among linked units denoted $\tau_{\mathcal{S}}$:} \begin{align} \label{SATE_tau_S}
    \tau_{\mathcal{S}} = \frac{1}{N_S} \sum_{\ell=1}^{N_S} Y_{\ell}(1) - Y_{\ell}(0).
\end{align}

\section{Implications for effect identifiability} \label{sec identifiability}

\KR{Record linkage affects causal identifiability through linkage errors and a covariate-dependent selection. Primarily, it interacts with the standard identifiability conditions for internal validity \citep{imbens_causal_2015, whatif, li_bayci_2023}. Falsely linked records corrupt consistency and missed links endanger positivity in finite samples while conditional exchangeability is not compromised. Ultimately, its preferential selection for distinctive profiles imposes to control for the shifted confounders between the baseline population $\mathcal A$ and the linked sample $\mathcal S$ for external validity. These threats, summarised in \Cref{tab:identifiability}, govern the two main components of the bias we develop later in \Cref{sec bias decompo}:} $$ \mathbb E [\widehat{\tau}_{\mathcal{S}}] - \tau_{\mathcal{A}} = \underbrace{\mathbb E [\widehat{\tau}_{\mathcal{S}}] - \tau_{\mathcal{S}}}_{\substack{\vphantom{g} \text{attenuation\vphantom{g} bias}\\ \text{from inconsistencies}}} + \underbrace{\vphantom{[]}\tau_{\mathcal{S}} - \tau_{\mathcal{A}}}_{\substack{\text{sampling bias}\\\text{from covariate shift}}}. $$

\begin{table}[ht]
\centering
\begin{tabularx}{\textwidth}{@{}XXXXX@{}}
\toprule
Validity & Condition & Threat & Bias & Solutions \\
\midrule
\multirow{9}{*}{Internal} & Consistency & Falsely linked records & Attenuation & FDP vanishes under \Cref{ass fdp} or through $\boldsymbol{Z}=\boldsymbol{X}$ (\Cref{rmk link var valid}) \\
\cmidrule{2-5} 
& Positivity & Missed links (finite sample only) & Extrapolation & Preferential selection protects atypical profiles under \Cref{ass random internal lack pos} \\
\cmidrule{2-5} 
& Conditional exchangeability & None &  &  \\
\midrule
\multirow{5}{*}{External} & \multirow{5}{*}{$S$-admissibility} & \Cref{S-admi} condition $(ii)$, no access to $\boldsymbol{X}_j$ & Selection & Control of $\boldsymbol{X}_j$ under \Cref{ass fdp} or through \Cref{rmk link var valid} \\
\cmidrule{3-5}
 &  & Covariate shift between $\mathcal{A}$ and $\mathcal{S}$ & Sample & Transport formula under \Cref{ass random internal lack pos}\\
\bottomrule
\end{tabularx}
\caption{\KR{Impact of record linkage on causal identification.}}
\label{tab:identifiability}
\end{table}

\subsection{Consistency} \label{sec consistency}

Consistency ensures that the observed outcome coincides with the potential outcome under the intervention: $$T=t \implies Y=Y(t) \text{, thus } Y = T \: Y(1) + (1-T) \: Y(0).$$ This entails that the treatment assignment mechanism does not affect the potential outcome response \citep{angrist_identification_1996, imbens_causal_2015}. \KR{We adopt the weaker distributional consistency \citep{epi_consistencyrelaxed, richardson2023distrconsistency, distrconsistencySCM24}, requiring only that the observed outcome be equivalent to the potential outcome in distribution. It suffices for the ATE we target and is better suited to linkage, where the observed outcome may belong to a different unit than the one whose potential outcome we target.}

\KR{The violation of consistency occurs when linked units potential outcomes do not correspond.} Indeed, recovering the ATE \KR{we would have obtained had we intervened and had we correctly linked the data} requires the linked units potential outcomes to share the same distribution. \KR{This is plausible if} \Cref{key ass hold} holds and if $T_i=T_j=t, \boldsymbol{X}_i=\boldsymbol{X}_j=\boldsymbol{x}$ since then $Y_i = Y_i(\boldsymbol{x},t) \sim Y_j(\boldsymbol{x},t) = Y_j$. Thereby, the outcome observed in $\mathcal{B}$ \KR{would reflect} the underlying potential outcome from $\mathcal{A}$, so distributional consistency, $Y_j \sim Y_i(\boldsymbol{x},t)$, \KR{would transfer} across the linkage. \KR{When records belong to distinct individuals who may differ in treatment group and covariates, this correspondence is disrupted since the units likely have different potential outcomes, breaking consistency.}

\KR{To preview how the violation materialises, consider treatment mismatch on the conditional outcome in the treated.} Let $T_j$ be the original treatment indicator and $T_i$ the treatment indicator after linkage of $i$ and $j$. By \Cref{key ass hold} the observed outcome is $Y_j=Y_j(T_j)$, meaning $$\left\{ \begin{array}{ll} Y_j=Y_j(T_i) \sim Y_i(T_i) \text{ under correct linkage}, \\ Y_j=Y_j(1-T_i) \sim Y_i(1-T_i) \text{ otherwise}. \end{array} \right.$$ The average conditional outcome in the treated is then: \begin{align*}
    \mathbb E [Y_j|T_i=1,\boldsymbol{X}] & = \mathbb E [Y_j|T_i=1,T_j=1,\boldsymbol{X}] \cdot \mathbb P (T_i=T_j|T_i=1,\boldsymbol{X}) \\ & \qquad \qquad + \mathbb E [Y_j|T_i=1,T_j=0,\boldsymbol{X}] \cdot \mathbb P (T_i \neq T_j|T_i=1,\boldsymbol{X}) \\
    & = \mathbb E [Y_j(1)|\boldsymbol{X}] \cdot \mathbb P (T_i=T_j|T_i=1,\boldsymbol{X}) + \mathbb E [Y_j(0)|\boldsymbol{X}] \cdot \mathbb P (T_i \neq T_j|T_i=1,\boldsymbol{X}) \\
    & = \mathbb E [Y_i(1)|\boldsymbol{X}] \cdot \mathbb P (T_i=T_j|T_i=1,\boldsymbol{X}) + \mathbb E [Y_i(0)|\boldsymbol{X}] \cdot \mathbb P (T_i \neq T_j|T_i=1,\boldsymbol{X}),
\end{align*} so distributional consistency only holds for the share of linked pairs with $T_i=T_j$.

\KR{Attenuation bias results from the treatment and covariate inconsistencies and takes different forms depending on the estimator used. Treatment mismatch affects all estimators, since observations contribute to inference on the opposite group \citep{wortman_RL_causality_2018}. Covariate mismatch affects estimators that condition on covariates.} For instance, the difference in means targets $f_Y(\boldsymbol{X}_j, t, \varepsilon_{Y_j})$, comparing $Y_j$ across levels of $T_i$; it is exposed to treatment mismatch. The outcome regression targets $f_Y(\boldsymbol{x}, t, \varepsilon_{Y_j})$, evaluating $\widehat\mu_{T_i=t}(\boldsymbol{X}_i=\boldsymbol{x})$; it is exposed to treatment and covariate mismatch. \KR{While inconsistencies threaten causal identification on linked data, their impact can be quantified, as we show in \Cref{sec bias decompo} and further in Appendix \Cref{sec attenuation bias}.}

\KR{Consistency can therefore only be restored through conditions on the linkage, providing a way to derive theoretical guarantees. An ideal case is when false discoveries vanish, achievable by raising the linkage decision threshold $\lambda$, albeit at the cost of more missed links.} \begin{assumption} \label{ass fdp} $\mathbb E[\text{FDP}] \underset{N_S \to \infty}{\longrightarrow} 0$. \end{assumption} An alternative is to link on all confounders and tolerate residual error, not always feasible in real applications. \begin{remark} \label{rmk link var valid} When the pre-treatment linking variables $\boldsymbol{Z}$ form a valid finite sample adjustment set, $\boldsymbol{X}=\boldsymbol{Z}$ are observed in $\mathcal{A}$ and in $\mathcal{B}$ and linkage performs `matching'. Since $T$ descends from $\boldsymbol{X}$, linking on informative covariates raises the probability that $\boldsymbol{X}_i, \boldsymbol{X}_j$ and $T_i, T_j$ coincide, yielding more accurate links, hence supporting distributional consistency. \end{remark} \KR{In many practical settings, it suffices to obtain a causal estimate up to a bounded error, which \Cref{sec bias decompo} quantifies.} 

\subsection{Positivity} \label{sec positivity}

Positivity requires that the probability of receiving any value of treatment conditional on covariates is bounded away from 0 and 1, essential for the joint distribution over all treatment values and all covariate profiles to be well defined: $$0 < \mathbb P (T=t|\boldsymbol{X}=\boldsymbol{x}) < 1 \text{ for all } t \in \{0,1\} \text{ and for all } \boldsymbol{x}.$$ 

\KR{Positivity transfers from the source population to the linked data under Assumptions \ref{ass linkage random} and \ref{key ass hold}. Indeed,} the selection mechanism is d-separated from the treatment by the covariates: $\mathbb P_{\mathcal{S}} (T=t|\boldsymbol{X}=\boldsymbol{x}) = \mathbb P_{\mathcal{A}} (T=t|\boldsymbol{X}=\boldsymbol{x},S=1) = \mathbb P_{\mathcal{A}} (T=t|\boldsymbol{X}=\boldsymbol{x}).$ Note that identification remains restricted to the covariate and treatment combinations supported in the observed data. For example, if only certain ethnicities participate from baseline to follow-up, or if only younger participants receive treatment and are followed up, positivity is foregone from the outset and biases emerge \citep{reflexRL_2020}. 

\KR{In finite samples however, record linkage causes the sample to shrink,} hampering the joint conditioning on treatment and confounders in $\mathcal{S}$ needed to transport the effect towards $\mathcal{A}$: $\mathbb P_{\mathcal{S}} (T=t,\boldsymbol{X}=\boldsymbol{x}) = \mathbb P_{\mathcal{S}} (T=t|\boldsymbol{X}=\boldsymbol{x}) \mathbb P_{\mathcal{S}}(\boldsymbol{X}=\boldsymbol{x})$. The data may suffer random non-positivity: $\mathbb P_{\mathcal{S}} (T=t|\boldsymbol{X}=\boldsymbol{x})=0$ for some $t \in \{0,1\}$ and some $\boldsymbol{x}$. The covariate space depletion may limit the overlap of supports and induce a covariate shift between distributions of $\mathcal{S}$ and $\mathcal{A}$: $\mathbb P_{\mathcal{S}}(\boldsymbol{X}=\boldsymbol{x}) \neq \mathbb P_{\mathcal{A}}(\boldsymbol{X}=\boldsymbol{x})$ for some $\boldsymbol{x}$. For example, our analysis may exclude untreated `Hispanic' women (non-positivity) or `White' women aged 30–35 from a population spanning 25–45 (limited overlap), these profiles being too common to be confidently linked. These empirical limitations primarily stem from missed links and worsen with the linkage decision rule stringency.

\KR{These gaps are benign in theory because linkage selection happens preferentially in the frail areas of the covariate space, which protects atypical profiles more at risk of lacking positivity. Methods for generalised treatment effects bridge these gaps by interpolation to transport the causal contrast from $\mathcal{S}$ to $\mathcal{A}$ at the cost of untestable modelling assumptions. Under suitable structure on the region of positivity and overlap violations, extrapolation bias can be avoided:} $R_{N_{S}} = \{ \boldsymbol{x} \in \boldsymbol{\mathcal{X}};\, \mathbb P_{\mathcal{A}}(\boldsymbol{X}=\boldsymbol{x})>0,\, \forall t \in \{0,1\}\,\mathbb P_{\mathcal{A}}(T=t|\boldsymbol{X}=\boldsymbol{x}) \in (0,1) \text{ and } \min( N_{S}(\boldsymbol{x}), N_{S0}(\boldsymbol{x}), N_{S1}(\boldsymbol{x}) )=0 \}$. \begin{assumption} \label{ass random internal lack pos} $R_{N_{S}} \subset \text{int }\boldsymbol{\mathcal{X}}$ and $\mathbb P (R_{N_{S}}) \underset{N_{S} \rightarrow \infty}{\longrightarrow} 0$, where $\text{int }\boldsymbol{\mathcal{X}}$ refers to the interior of the covariate space. \end{assumption} That is, violations are finite sample artefacts, internal to the covariate space, supporting transportability and estimation via interpolation \citep{com_positivity, whatif, comp_gmethogs_randpos}.

\subsection{Conditional exchangeability} \label{sec exchangeability} 

The conditional distribution of the potential outcomes under treatment $t \in \{0,1\}$ is the same among both treatment groups $\{T=0\}$ and $\{T=1\}$ given $\boldsymbol{X}=\boldsymbol{x}$, \citep{imbens_causal_2015, whatif}: $$Y(t) \indep T \mid \boldsymbol{X} \text{ for } t \in \{0,1\}.$$ This implies mean exchangeability over treatment groups, $\mathbb E [Y(t)|\boldsymbol{X}=\boldsymbol{x}] = \mathbb E [Y(t)|T=1, \boldsymbol{X}=\boldsymbol{x}] = \mathbb E [Y(t)|T=0, \boldsymbol{X}=\boldsymbol{x}]$. In other terms, this condition states that there is no unmeasured confounding. 

Linkage does not interfere with conditional exchangeability. Under \Cref{key ass hold}, the counterfactuals are independent of treatment assignment given the observed covariates in the source population. This condition is directly encoded in \Cref{subfig:Event1} since $Y(t)$ and $T$ are d-separated by $\boldsymbol{X}$. With record linkage, linked pairs become the exchangeable units at stake, and this conditional independence extends to both correctly and incorrectly linked units. Indeed $T_i$ and $Y_j(t)$ are d-separated by $\{\boldsymbol{X}_i, S_{i,j}\}$ since the only path connecting them in \Cref{subfig:Event2} is blocked: it contains a non-collider $\boldsymbol{X}_i$ that is conditioned on \citep{mit_causal_2001, pearl_2009}. 

Intuitively, suppose that, given BMI, hormone levels, cardiovascular measures, medication, and lifestyle, counterfactual bone densities do not depend on whether individuals are assigned to exercise or not. Then, if we were to randomly relabel who is assigned to exercise in the population, the potential outcomes $Y(0), Y(1)$ of the relabelled individuals---what would have happened under either exercise regime---would remain unchanged. This relabelling thought experiment extends to linked data: the potential outcomes of individuals in $\mathcal{B}$ linked to individuals in $\mathcal{A}$ are not affected by which treatment label they end up matched to. 

\subsection{$S$-admissibility} \label{sec recoverability}

\KR{The effect in the source population is recoverable from the linked sample as long as the follow-up covariates $\boldsymbol X_j$ can be controlled. The conditions above discussed the internal validity of $\mathcal S$. Moreover, the external validity is at stake because the covariate distribution in $\mathcal S$ differs from that in $\mathcal A$. Consequently, identifying $\tau_{\mathcal A}$ requires not only that the common causal assumptions hold in $\mathcal S$, but also that the effect be recoverable from selection bias.} We thus examine the causal record linkage problem through the lens of data fusion, where the key challenge is to recover $\tau_{\mathcal A}$ despite restricting inference to reliably linked pairs.

Causal identification under selection relies on the `selection backdoor criterion' \citep{BareinboimTianPearl2014, bareinboimpearl2016, Kenah_2023}. \begin{definition}[selection backdoor criterion] \label{S-admi} Given a causal system augmented with the sampling mechanism $S$, where treatment $T$ has no other descendant than outcome $Y$, a set of confounders $\boldsymbol{X}$ satisfies the selection backdoor criterion if \begin{itemize}
    \item[(i)] $Y(t) \indep T | \boldsymbol{X}$,
    \item[(ii)] $Y(t) \indep S | \boldsymbol{X}, T$,
    \item[(iii)] $\{\boldsymbol{X}, T, Y\}$ are observed in the selection biased data such that $S=1$,
    \item[(iv)] $\boldsymbol{X}$ is observed at the population level.
\end{itemize} \end{definition} We say, equivalently, that there is no unmeasured sample selection bias \citep{Kenah_2023}; that $\boldsymbol{X}$ is $S$-backdoor admissible (or $S$-admissible) \citep{pearlbareinboim2014}; that $p(Y(t))$ is $S$-recoverable from selection biased data i.e.\ expressible in terms of the distribution under selection $S$ \citep{BareinboimTianPearl2014}; or that sample selection bias can be corrected \citep{ipsw_2004}.

In the causal record linkage system of \Cref{subfig:Event2}, condition $(ii)$ from \Cref{S-admi} fails without further assumptions, as $Y_j(t)\not\!\!\!\indep S_{i,j} | \boldsymbol{X}_i, T_i$; conditioning on $\boldsymbol{X}_j$ would instead satisfy it. $\boldsymbol{X}_j$ is accessible either when the FDP vanishes, as linked units then refer to the same individuals giving $\boldsymbol{X}_i=\boldsymbol{X}_j$ (under \Cref{ass fdp} up to a small finite sample error) or when all covariates are linking variables, so $\boldsymbol{Z}_j = \boldsymbol{X}_j$ is observed (\Cref{rmk link var valid}). In practice, inference should be treated with caution, or abandoned, when the FDP is not controlled. Further, the point of \Cref{rmk link var valid} requires careful discussion in application, although in finite samples, valid and asymptotically optimal adjustment sets may differ \citep{rutsch2025finitesampleoptimaladjustmentsets}, and $\boldsymbol{Z}$ may proxy $\boldsymbol{X}$. 

\KR{Beyond condition $(ii)$, the conditions from \Cref{S-admi} hold by definition of the causal record linkage task, guaranteeing $S$-recoverability of the effect under control over $\boldsymbol{X}_j$,} but not its estimation in practice: the linkage selection \KR{shrinks the covariate space relative to $\mathcal{A}$, inducing the covariate shift,} and finite sample artefacts may induce positivity and overlap violations. These concerns remain manageable, as they occur at random and internally to the covariate space, as formalised by \Cref{ass random internal lack pos}, which supports estimation of the transport formula between the selected data and the baseline.

\begin{theorem} \label{result_1} In the causal record linkage context represented in \Cref{fig:SelectionDiagram}, under assumptions \ref{ass linkage random}, \ref{key ass hold}, \ref{ass fdp}, \ref{ass random internal lack pos}, $\tau_{\mathcal{A}} = \mathbb E_{\mathcal{A}}[\tau_{\mathcal{S}}(\boldsymbol{X})]$ is identified and estimable from the available data. \end{theorem} 

\KR{In the above, $\tau_{\mathcal{A}}$ is the estimand of \Cref{estimand_tau_A} and $\tau_{\mathcal{S}}$ is the sample ATE of \Cref{SATE_tau_S}; $\tau_{\mathcal{S}}(\boldsymbol{X})$ is the sample conditional ATE in the selected linked units.} The proof of \Cref{result_1} is deferred to the Appendix \Cref{appendix proof}. \KR{Note that \Cref{ass fdp} on the FDP is met by stricter linkage while \Cref{ass random internal lack pos} needs the linked sample to be as large as possible; both may not hold in practice but the error may be negligible, especially when more covariates are linking variables---see the earlier discussion in \Cref{sec consistency}. }

\KR{To conclude, ensuring identification requires imposing stringency on the linkage, trading consistency, and correspondingly attenuation bias, against the covariate shift and its sampling bias. If the FDP vanishes (\Cref{ass fdp}), consistency and $S$-admissibility are restored, and the covariate shift is corrected by the transport formula} under \Cref{ass random internal lack pos} so $\tau_{\mathcal A}$ is recoverable. \Cref{sec bias decompo} quantifies the bias trade-off, and \Cref{sec estimation} describes the tools available for estimating the transport formula. The simulations of \Cref{sec simulations} will later let us observe concretely how these biases interplay with the effect estimation.

\section{Bias decomposition} \label{sec bias decompo}

Record linkage may be interpreted as a matching procedure for estimating treatment effects from multiple data sources. Like a nearest neighbour estimator, it links each baseline unit to a follow-up unit with similar values of pre-treatment linking variables $\boldsymbol{Z} \subseteq \boldsymbol{X}$, thereby approximating the potential outcome $Y_i(t)$ of a unit $i$ with the observed outcome $Y_j | T_j = t$ of a potentially distinct unit $j$. In this Section, we examine the implications of this procedure, contrasting estimators built on linked data to the targeted estimand. Especially, we quantify in \Cref{result_2} how falsely linked data bias causal estimates.

We distinguish different standard consistent ATE estimators: difference in means, propensity score stratification, inverse probability weighting, matching and outcome regression (compared in the application and simulations Sections \ref{sec application} and \ref{sec simulations}, and developed in Appendix \Cref{appendix ate}); from non-adjusted methods, through indirect adjustment via the propensity score, to outcome regression controlling for covariates. This section focuses, as a simplified case, on the bias of the difference in means estimator relative to the average causal contrast in the baseline population. Such estimator may be seen as an outcome regression estimator with omitted variables \citep{causal_error_estimators_26}, and while extending this analysis formally to the remaining estimators is left out for future work, we hope this development helps clarify the biases at stake when estimating causal effects from linked data.

The difference in means ATE estimator computed on the selected linked data is described as \begin{equation} \label{estimator_hat_tau_S} \widehat{\tau}_{\mathcal{S}} = \frac{1}{N_{S1}} \sum_{i,j} S_{i,j} \, T_i \, Y_j - \frac{1}{N_{S0}} \sum_{i,j} S_{i,j} \, (1-T_i) \, Y_j.\end{equation} We compare $\widehat{\tau}_{\mathcal{S}}$ with our target, the ATE in the baseline population from \Cref{estimand_tau_A}, to study the bias of the causal record linkage procedure: $$\mathbb E [\widehat{\tau}_{\mathcal{S}}] - \tau_{\mathcal{A}} = \mathbb E [ \widehat{\zeta}_{N_{S}} ] + \mathbb E [ \widehat{\xi}_{N_{S}} ],$$ with $\widehat{\zeta}_{N_{S}}$ the sampling bias and $\widehat{\xi}_{N_{S}}$ the attenuation bias, defined as follows: \begin{align*}
    \widehat{\zeta}_{N_{S}} = & \frac{1}{N_{S1}} \sum_{i,j} \big\{ S_{i,j} \, T_i \, \mu_1(\boldsymbol{X}_i) \big\} - \frac{1}{N_{S0}} \sum_{i,j} \big\{ S_{i,j} \, (1-T_i) \, \mu_0(\boldsymbol{X}_i) \big\} - \tau_{\mathcal{A}},\\
    \widehat{\xi}_{N_{S}} = & \frac{1}{N_{S1}} \sum_{i,j} \big\{ S_{i,j} \,T_i \, (Y_j - \mu_1(\boldsymbol{X}_j)) \big\} - \frac{1}{N_{S0}} \sum_{i,j} \big\{ S_{i,j} \, (1-T_i) \, (Y_j - \mu_0(\boldsymbol{X}_j)) \big\} \\
    & + \frac{1}{N_{S0}} \sum_{i,j} \big\{ S_{i,j} \, (1-T_i) \, (\mu_0(\boldsymbol{X}_i)-\mu_0(\boldsymbol{X}_j)) \big\}  - \frac{1}{N_{S1}} \sum_{i,j} \big\{ S_{i,j} \, T_i \, (\mu_1(\boldsymbol{X}_i)-\mu_1(\boldsymbol{X}_j))\big\}.
\end{align*}

The sampling bias captures the difference in ATE between the selected linked sample $\mathcal{S}$ and the baseline population $\mathcal{A}$ due to missed links (FN). It results from \KR{the covariate shift (brought up in \Cref{sec positivity} and developed in \Cref{sec recoverability}) between distributions in $\mathcal{S}$ and in $\mathcal{A}$} intensifying with linkage stringency. 

The attenuation bias (announced in \Cref{sec consistency}) is due to falsely linked records (FP) and emerges via two mechanisms. The first mechanism implies a contradiction in the causal effect from mismatches in treatment groups: $Y_j|T_j=1-t,\boldsymbol{X}_j$ contrasted with $\mu_t(\boldsymbol{X}_j)$. The magnitude of this part of the bias is the opposite of the causal effect targeted and it appears in proportion to the share of FP pairing units from different treatment groups \citep{wortman_rl_causal}.

The second mechanism arises from discrepancies between covariates inducing differences in linked units potential outcomes: $\mu_t(\boldsymbol{X}_j)$ contrasted with $\mu_t(\boldsymbol{X}_i)$. This part of the bias is analogous to the conditional bias of matching estimators due to matching discrepancy \citep{AandIMatching2006, AandIMatching2011, AandIMatching2012, AandIMatching2016}. It is a manifestation of coefficient dilution arising from errors in independent variables \citep{dilution_spearman, ml_noise_shrink, dilution_thompson, Tancredi_Liseo_linked_2015, Wang_Slawski_linked_2022}.

The impact of consistency violations on estimation, that takes shape as attenuation bias, depends on the ATE estimator used, and on whether the potential outcome it targets is defined at the level of the covariates or in its aggregated form, as discussed in \Cref{sec consistency}. To frame this impact in the context of the difference in means estimator, we introduce additional assumptions.

When using record linkage for causal inference, we presume to be able to recover a suitable sample. Although all observations are assumed eligible for treatment, not all profiles are being selected. We therefore require that record linkage recover a positive fraction of both treated and control observations. \begin{assumption} \label{ass treated control} $\forall t \in \{0,1\},\, \mathbb E [N_{St}]>0$. \end{assumption} Alternatively we could link treated and untreated subsets from $\mathcal{A}$ to $\mathcal{B}$ separately and discard pairs involving records in $\mathcal{B}$ linked to both treatment groups. 

We also require that linkage errors do not systematically confound treatment groups ($T_i=t, T_j=1-t$), i.e.\ the majority of falsely linked pairs should connect entities who received the same treatment ($T_i=t, T_j=t$). \begin{assumption} \label{ass most fp} $\forall t \in \{0,1\},\, \mathbb E [N_{S \nabla T_i=t, T_j=1-t}] \leq \mathbb E [N_{S \nabla T_i=t, T_j=t}]$. \end{assumption}

Finally we need a consistent estimation method. \begin{assumption} \label{ass consisctency ate} The ATE estimator used is consistent. \end{assumption} The difference in means estimator is consistent for the linked sample ATE when the treatment is randomly assigned. Other estimators may additionally require correctly specified models for the propensity score or for the outcome, for instance. \begin{theorem} \label{result_2} In the causal record linkage context represented in \Cref{fig:SelectionDiagram}, under assumptions \ref{ass linkage random}, \ref{key ass hold},  \ref{ass delta separated}, \ref{ass treated control}, \ref{ass most fp}, \ref{ass consisctency ate}, the causal effect estimation suffers from attenuation bias due to falsely linked data: $\mathbb E [\widehat{\tau}_{\mathcal{S}}] = \tau_{\mathcal{S}} (1 - \mathbb E [\text{FDP}]).$ \end{theorem}

With Theorems \ref{result_1} and \ref{result_2}, we have characterised the gap between the estimator and the estimand in two steps. The ATE estimator computed on linked data, although converging towards the true ATE on the selected sample, is hampered by attenuation bias (\Cref{result_2}). This ATE on the selected sample differs from the targeted ATE on the baseline population due to sampling bias induced by selection (\Cref{result_1}). For instance, if individuals with extreme BMI are more confidently linked, the effect of physical activity on bone density in the selected sample may differ from the targeted effect over the whole baseline population, which spans a broader range of BMI values. As a result, recovering the causal effect in the baseline population requires eliminating attenuation bias and transporting the effect over samples. 

Consider the implications of these results. \KR{The trade-off between inconsistencies and covariate shift in the causal identification} transposes into a trade-off between attenuation and sampling bias in the inference: a balance between falsely linked records and missed links, FDP and sensitivity, bias and variance. If the linked data are mostly true links, sampling bias drives our estimator. If falsely linked data link records from different treatment groups, our estimator accounts for adversarial contributions to the causal effect and, if they link records within the same treatment group, our estimator behaves like a one-to-one matching estimator; the main concern in that case becomes `how similar are the potential outcomes of the falsely linked records?'

Attenuation bias dominates when the FDP is non-negligible, while sampling bias dominates when the decision rule is too strict. When the linked sample is large enough to support causal inference, this suggests favouring a relatively stringent decision rule so as to render attenuation bias tolerable, while the resulting sampling bias could still be corrected for under the assumptions that make \Cref{S-admi} valid. Note that, were the population unchanged from baseline to follow-up ($N_{\mathcal{A}}=N_{\mathcal{B}}=\sum_{i,j} \Delta_{i,j}$), the record linkage task would be easier \citep{fastLink_paper_2019, review_methods_2014, theoryRL_2018, robachfdp}, the expected potential outcomes in $\mathcal{A}$ and $\mathcal{B}$ would coincide on average so the bias would only stem from opposite contributions, which would also be less likely.

\section{Estimation} \label{sec estimation}

\KR{The remaining obstacle to estimating $\tau_{\mathcal A}$ on linked data is the covariate shift between $\mathcal S$ and $\mathcal A$, which is handled at the estimation stage. Unbiased causal effect estimation on linked data hinges on a stringent record linkage decision rule, which ensures consistency and effect recoverability from the selected data $\mathcal S$ (\Cref{sec identifiability}). Linkage thereby favours atypical profiles and the set $\mathcal S$ of observations from $\mathcal A$ augmented with outcome information from $\mathcal B$ is not representative. Even with negligible falsely linked records, a standard estimator applied to $\mathcal S$ therefore targets $\tau_{\mathcal S}$ rather than $\tau_{\mathcal A}$ (\Cref{sec bias decompo}). We present three families of generalised treatment effect (g-)methods that correct this shift.}

G-methods cope with this issue by exploiting the representativeness of $\mathcal{A}$, at the cost of untestable interpolation \citep{whatif, dahabreh_transport, colnetetal_2024}. We consider three families of estimators, used in the simulations of \Cref{sec simulations} and in the real data application of \Cref{sec application}. The g-formula \citep{Robins_1986_g} models conditional outcomes and marginalises over the targeted population. Weighting \citep{stuart_ipw, cole_g, Hainmueller_2012} uses propensity score modelling to transform observations and reflect the targeted population characteristics. Doubly robust approaches combine both and remain consistent if either model is correctly specified.

The two ingredients for reliable causal effect estimation on linked data are formalised with Theorems \ref{result_1} and \ref{result_2}. The generalised treatment effect estimators can correct the sampling bias and recover $\tau_{\mathcal{A}}$ from $\mathcal{S}$ despite the selection induced by linkage. Though, they cannot correct the attenuation bias due to falsely linked records. Nevertheless, when linkage is imperfect, matching similar profiles remains adequate to identify causal effects, or at least their direction as we illustrate in \Cref{sec application}. The intuition being that falsely linked individual records may still create an instance for which the observed outcome is a reasonable potential outcome.

\subsection{G-formula estimators} \label{sec g formula}

G-formula estimators use outcome modelling to generalise findings from observations at hand towards the targeted population \citep{Robins_1986_g, dahabreh2019, dahabreh_transport, colnetetal_2024}. To wit, the potential outcome distributions are learnt from the selected linked data and applied on the baseline population. The classical g-formula treatment effect estimator $\widehat{\tau}_{\text{G}}$ is defined as \begin{equation} \label{tau_g_formula}
     \frac{1}{\nA} \sum_{i=1}^{\nA} \widehat{\mu}_{1,\mathcal{S}}(\boldsymbol{x}_i) - \widehat{\mu}_{0,\mathcal{S}}(\boldsymbol{x}_i),
\end{equation} where $\widehat{\mu}_{t,\mathcal{S}}(\boldsymbol{x})$ is fitted using the selected linked data for treatment group $t$. When \Cref{ass random internal lack pos} necessary to generalise the effect from $\mathcal{S}$ to $\mathcal{A}$ is not reasonable, causal identification is confined to regions where positivity holds. We can correct this estimator with a Prediction Powered Inference (PPI) term accounting for the difference between modelled potential and observed outcomes \citep{gronsbell2025ppi}. The use of such estimator for generalised causal contrast is fairly novel \citep{PPCI25}.  We define the PPI g-formula estimator $\widehat\tau_{ \text{G}_{ \text{PPI} } }$ as: \begin{equation} \label{tau_g_formula_ppi}
     \frac{1}{\nA} \sum_{i=1}^{\nA} \widehat\omega_i \, (\widehat\mu_{1,\mathcal{S}}(\boldsymbol{x}_i) - \widehat\mu_{0,\mathcal{S}}(\boldsymbol{x}_i)) + \frac{1}{N_{S}} \sum_{\ell=1}^{N_{S}} t_\ell \, ( y_\ell - \widehat\omega_\ell \cdot \widehat\mu_{1,\mathcal{S}}(\boldsymbol{x}_\ell) ) - (1 - t_\ell) \, ( y_\ell - \widehat\omega_\ell \cdot \widehat\mu_{0,\mathcal{S}}(\boldsymbol{x}_\ell) ) .
\end{equation} Individual weights $\widehat\omega_i$ are included to obtain a more efficient estimator, with lower variance; they control the contribution of the estimated potential outcomes into estimation, adapting to the accuracy between potential and observed outcomes of the predictive model used---see \citep{angelopoulos2024, gronsbell2025ppi, ipd_package} for further details.

\subsection{Weighting estimators} \label{sec weighting}

Weighting estimators calibrate the observations from the selected linked data to represent the baseline population. They transport evidence from one population to the other by accounting for the covariate distribution shift between populations with weights $\omega$. The evidence in the selected linked population is derived from inverse probability weighting using propensity score estimation. As a consequence of their definition, weighting estimators might be sensitive to the lack of random positivity \citep{epidm_2000, weight_positivity_2008, whatif}. The expected conditional outcome in treatment group $t$ is identified as: $$\mathbb E_{\mathcal{A}} [\, \mathbb E_{\mathcal{S}} [Y | T=t, \boldsymbol{X}=\boldsymbol{x}]\, ] = 
\mathbb E_{\mathcal{S}} [\, \mathbb E_{\mathcal{S}} [Y | T=t, \boldsymbol{X}=\boldsymbol{x}] \, \omega\, ] = \mathbb E_{\mathcal{S}} \Big[\, \frac{\mathbb E_{\mathcal{S}} [Y \, \ind_{\{T=t\}} | \boldsymbol{X}=\boldsymbol{x}]}{\mathbb P_{\mathcal{S}} (T=t|\boldsymbol{X}=\boldsymbol{x})} \, \omega\, \Big].$$ We then define the IPSW estimator $\widehat\tau_{\text{IPSW}}$ and the EB estimator $\widehat\tau_{\text{EB}}$ for the treatment effect with: \begin{equation} \label{tau_weighting}
    \frac{1}{N_{S}} \sum_{\ell=1}^{N_{S}} \widehat\omega_\ell \, y_\ell \, \Big( \frac{t_\ell}{\widehat\pi(\boldsymbol{x}_\ell)} - \frac{1-t_\ell}{1-\widehat\pi(\boldsymbol{x}_\ell)} \Big).
\end{equation} 

These estimators differ in their weighting scheme. Entropy Balancing (EB) or Calibration Weighting (CW) \citep{Hainmueller_2012, colnetetal_2024, WeightIt} constructs balancing weights by minimising the negative entropy subject to balancing and normalising constraints, aligning the moments of the selected sample $\mathcal{S}$ with those of the baseline population $\mathcal{A}$. Inverse Probability of Sampling Weighting (IPSW) \citep{stuart_ipw, cole_g, colnetetal_2024} instead weights observations by the covariate density ratio $\mathbb P_{\mathcal{A}}(\boldsymbol{X}=\boldsymbol{x})/\mathbb P_{\mathcal{S}}(\boldsymbol{X}=\boldsymbol{x})$ which can be developed into $\mathbb P_{\mathcal{A}}(S=1) / \mathbb P_{\mathcal{A}}(S=1|\boldsymbol{X})$ \citep{cole_g, colnetetal_2024}. We estimate these sampling weights using a logistic regression.

We estimate the propensity score with a logistic regression, on the baseline $\mathcal{A}$: $\pi_\mathcal{A}(\boldsymbol{X}) = \mathbb P_{\mathcal{A}}(T=1|\boldsymbol{X})$, and on the restricted selected subset $\mathcal{S}$: $\pi_{\mathcal{S}}(\boldsymbol{X}) = \mathbb P_{\mathcal{S}}(T=1|\boldsymbol{X})$. As noted in \Cref{sec positivity} these propensity scores should coincide: $\mathbb P_{\mathcal{S}} (T|\boldsymbol{X}) = \mathbb P_{\mathcal{A}} (T|\boldsymbol{X})$, although they tend to differ empirically. According to the original work, we should use $\pi_{\mathcal{S}}(\boldsymbol{X})$ as we generalise from population $\mathcal{S}$ \citep{stuart_ipw, cole_g, colnetetal_2024}. In the simulation study, using the propensity score estimated in the more stable baseline population yields better estimates. In the real data application, following the original idea yields more accurate estimates.

\subsection{Doubly robust estimators} \label{sec doubly robust}

Doubly robust versions of these estimators exist \citep{doublyrob, leeAEB, colnetetal_2024}. They combine modelling of the outcome \Cref{tau_g_formula} and of the treatment probability \Cref{tau_weighting} to form the Augmented IPSW $\widehat\tau_{\text{AIPSW}}$ and the Augmented EB $\widehat\tau_{\text{AEB}}$ estimators: \begin{equation} \label{tau_aug} \frac{1}{\nA} \sum_{i=1}^{\nA} \widehat\mu_{1,\mathcal{S}}(\boldsymbol{x}_i) - \widehat\mu_{0,\mathcal{S}}(\boldsymbol{x}_i) + \frac{1}{N_{S}} \sum_{\ell=1}^{N_{S}} \widehat\omega_\ell \Big( \frac{t_\ell \, (y_\ell - \widehat\mu_{1,\mathcal{S}}(\boldsymbol{x}_\ell))}{\widehat\pi(\boldsymbol{x}_\ell)} - \frac{(1-t_\ell) \, (y_\ell - \widehat\mu_{0,\mathcal{S}}(\boldsymbol{x}_\ell))}{1-\widehat\pi(\boldsymbol{x}_\ell)} \Big) .\end{equation} These estimators are consistent when one of the models (outcome or propensity score) is correctly specified.

\KR{All three families of estimators can address the covariate shift between $\mathcal{A}$ and $\mathcal{S}$ that gives rise to sampling bias. However, they require sufficiently large treatment arms in $\mathcal{S}$ to fit the outcome or propensity models. Moreover, the g-formula relies on a correctly specified outcome model and risks extrapolating beyond the linked support. Weighting estimators require overlap between the supports of $\mathcal{A}$ and $\mathcal{S}$ and can become unstable as these supports diverge. Doubly robust estimators combine both approaches and remain consistent if either model is correctly specified.}

\KR{As with the assumptions underlying \Cref{result_1}, these requirements are in tension with the linkage stringency needed to eliminate attenuation bias, which shrinks $\mathcal{S}$ and increases its divergence from $\mathcal{A}$. Estimator performance therefore depends on the application. Under substantial covariate shift, the simulations (\Cref{sec simulations}) show that weighting fails while outcome modelling recovers the effect. In contrast, in the application (\Cref{sec application}), all estimators exhibit biases of similar magnitude. The assumptions underlying each estimator thus inform the linkage decisions discussed in the next section.}

\section{Guiding linkage decisions} \label{sec linkage selection}

Reliable causal effect estimation with linked data is contingent on identifiability of the causal contrast in the linked data, recoverability of the experimental distribution from linkage sampling bias, and generalisability of the estimation towards the intended target of inference. The linked data selection thus calls for complementary evaluation measures to assess record linkage quality and validity of the causal framework. Having established the main condition under which the selection backdoor criterion holds: \Cref{ass fdp}, allowing us to recover the causal effect in theory, we now provide guidelines for jointly navigating record linkage and causal framework in order to estimate the causal effect in practice.

Before record linkage, it is worth gauging how hard the task is. A naive linkage approach---matching records on exact agreement of identifying information---reveals the discriminative power of the linking variables, indicating how much ambiguity the linkage procedure will face. 

\paragraph{Step 1---Record linkage across decision thresholds $\lambda$} Run the record linkage procedure and apply a grid of cutoffs $\lambda \in [0.5,1)$ to the potential pairs of records in the output. This will shed light on the trade-offs selection induces (linkage quality versus linked sample size versus representativeness of the baseline).

\paragraph{Step 2---Examine the linkage output} Record linkage performance may be estimated without unique identifiers \citep{belin_rubin_1995, sensitivity_2002, fdr_genomewide_2003, eval_linkage_17, fastLink_paper_2019, measurementError_2019, sensitivity_2020, DG_block_esterror, DG_esterror22, linkage_bias_datasynthesis24, robachfdp}; here we focus on controlling the estimated FDP. The default output ($\lambda=0.5$) gives the largest linked set and the most reliable FDP estimate. Estimating the FDP across thresholds from $0.5$ up to the highest linkage score observed shows the potential gains from stricter linkage (but FDP estimation is destabilised by smaller samples). The estimated FDP bounds the attenuation bias of the linked sample ATE estimator and helps assess \Cref{ass fdp}: controlling it ensures consistency in the linked data and validity of the selection backdoor criterion. One should also track the linked sample size, overall and within treatment subgroups, as a function of $\lambda$ to assess \Cref{ass treated control}, since stricter linkage trades sample size for linkage quality. What constitutes a reasonable FDP should be judged per application.

\paragraph{Step 3---Scrutinise the resulting linked data} The linked sample should remain faithful to the baseline population. Generalisability and effect estimation remain possible when the analytic support is smaller than that of the target population, provided the lack of positivity and overlap is at random rather than structural \citep{tipton_ps_2013, colnetetal_2024}, and violations are internal \citep{com_positivity, comp_gmethogs_randpos}. However, theory only guarantees identification in regions of the covariate space shared by the linked sample $\mathcal{S}$ and the baseline population $\mathcal{A}$, and the sampling bias resulting from an increasingly small and atypical sample of data linked confidently may dominate the estimator. Deviations between $\mathcal{S}$ and $\mathcal{A}$ may be assessed with the Maximum Mean Discrepancy (MMD) between covariate distributions, the Intersection over Union (IoU) between variable supports, the Standardised Mean Difference (SMD) between variables. In addition, \KR{the potential positivity and overlap violations in the linked sample $\mathcal{S}$ could be examined to assess \Cref{ass random internal lack pos}.} Covariate distributions could be compared across treatment status $\{T=1\}$ versus $\{T=0\}$ and samples $\mathcal{A}$ versus $\mathcal{S}$. 

\paragraph{Step 4---Decide whether to proceed with causal inference} Based on the joint behaviour of \KR{estimated FDP, deviations between $\mathcal{S}$ and $\mathcal{A}$ and sample size}, determine whether the linked data supports valid inference or whether inference should be abandoned. As a practical rule, restrict candidate cutoffs to those retaining a minimum viable sample per treatment arm (e.g.\ in the application of \Cref{sec application} we target at least 100 linked units per group), then select the strictest $\lambda$ at the `elbow' of the MMD curve---the point where linked and baseline samples begin to diverge---while monitoring $\widehat{\text{FDP}}$.

\paragraph{Step 5---Apply generalised estimators to the validated linked sample} Once the linked sample $\mathcal{S}$ is built, compute standard estimators and g-methods. Although the principle of separating design and analysis cautions against letting design choices inform estimation, comparing results across thresholds $\lambda$ may reveal the shifting balance between attenuation bias (from falsely linked pairs) and sampling bias (from missed links). This trade-off however, may be ambiguous as standard error has many roots and, in the causal record linkage setting where linking variables and covariates are intertwined, there is no evidence that increased variance, or deviations from the main effect, reflects linkage error.

\begin{figure}[!ht]
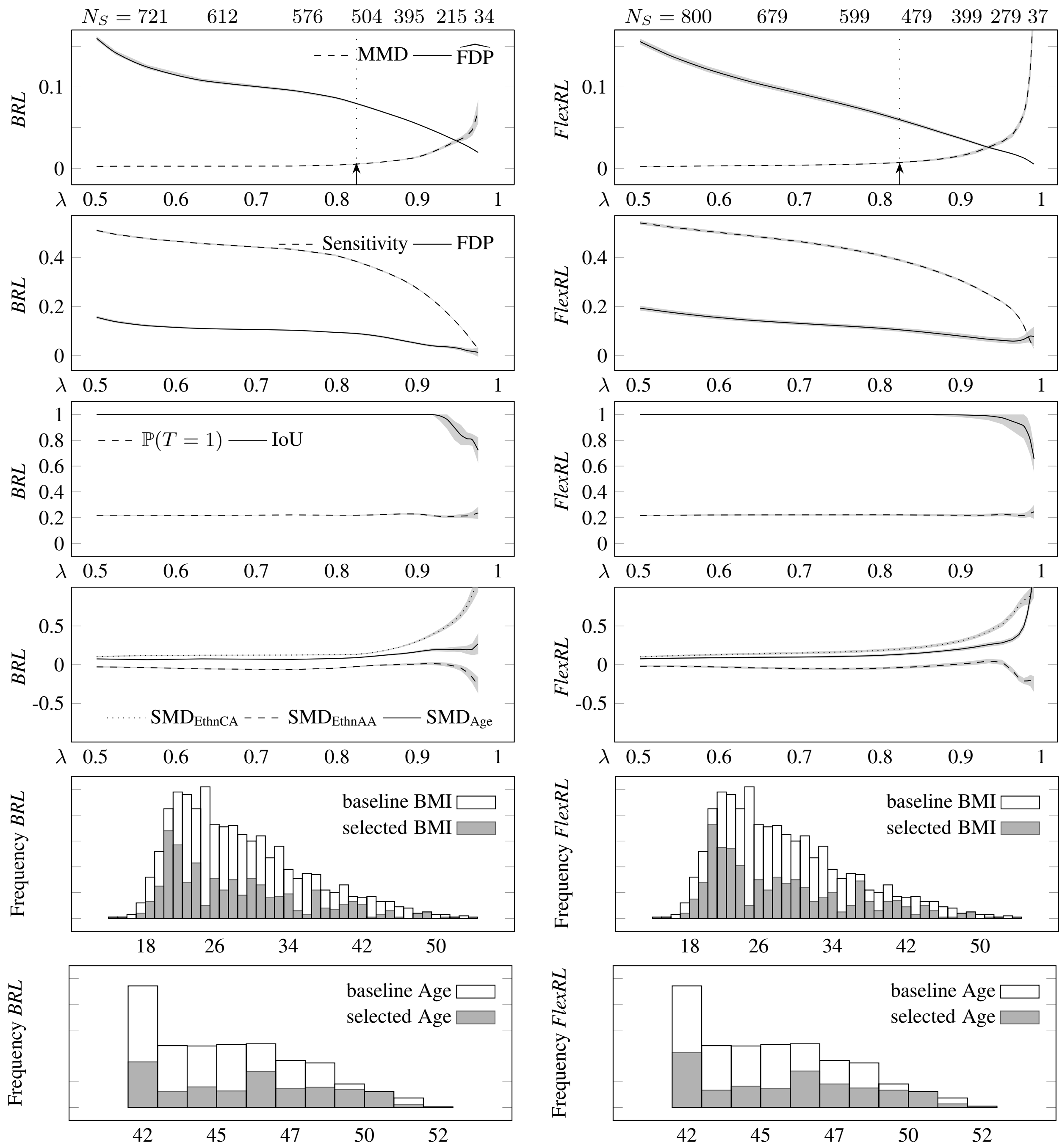

    \centering
    \input{Figures/mmd_fdthat}
    \input{Figures/sensitivity_fdp}
    \input{Figures/probatreat_iou}
    \input{Figures/smd}
    \begin{tikzpicture}
    % BRL BMI HIST
    \begin{axis}[
        width=8.5cm,  
        height=4cm,
        ymin=-10, ymax=110,
        ytick={0,20,40,60,80,100},
        yticklabels={\hphantom{0.4},,,,,},
        xtick style={opacity=0},
        xtick={17.5,25.5,33.5,41.5,49.5},
        xticklabels={\hphantom{1}18,26,34,42,50},
        ylabel={Frequency \textit{BRL}},
        ylabel style={yshift=-1.5pt},
        legend style={draw=none, anchor=south west, fill=none, at={(0.585,0.55)}},
        legend plot pos=right,
        legend entries={baseline\hphantom{tec}\!\!\!\!\!\!\! BMI, selected\hphantom{ain}\!\!\!\!\!\!\! BMI}
        ]
    \addplot[ybar interval, mark=no, draw=black, area legend] plot coordinates 
    { % baseline BMI
    (14, 1) (15, 1) (16, 3) (17, 12) (18, 32) (19, 52) (20, 85) (21, 98) (22, 96) (23, 85) (24, 102) (25, 73) (26, 71) (27, 72) (28, 65) (29, 60) (30, 64) (31, 47) (32, 56) (33, 38) (34, 31) (35, 35) (36, 34) (37, 22) (38, 20) (39, 26) (40, 17) (41, 16) (42, 17) (43, 13) (44, 13) (45, 9) (46, 8) (47, 4) (48, 5) (49, 3) (50, 3) (51, 1) (52, 2) (53, 1) (54, 1)
    };
    \addplot[ybar interval, mark=no, draw=black, fill=black!50, opacity=0.6, area legend] plot coordinates 
    { % linked BMI
    (14, 1) (15, 1) (16, 0) (17, 4) (18, 13) (19, 25) (20, 68) (21, 57) (22, 28) (23, 43) (24, 10) (25, 31) (26, 22) (27, 30) (28, 18) (29, 31) (30, 26) (31, 16) (32, 17) (33, 19) (34, 6) (35, 3) (36, 22) (37, 8) (38, 7) (39, 11) (40, 13) (41, 11) (42, 1) (43, 3) (44, 6) (45, 1) (46, 0) (47, 4) (48, 5) (49, 0) (50, 1) (51, 1) (52, 0) (53, 0) (54, 0)
    };\end{axis}
    \end{tikzpicture}
    \hspace{10pt}
    \begin{tikzpicture}
    % FLEXRL BMI
    \begin{axis}[
        width=8.5cm,  
        height=4cm,
        ymin=-10, ymax=110,
        ytick={0,20,40,60,80,100},
        yticklabels={\hphantom{0.4},,,,,},
        xtick style={opacity=0},
        xtick={17.5,25.5,33.5,41.5,49.5},
        xticklabels={\hphantom{1}18,26,34,42,50},
        ylabel={Frequency \textit{FlexRL}},
        ylabel style={yshift=-1.5pt},
        legend style={draw=none, anchor=south west, fill=none, at={(0.585,0.55)}},
        legend plot pos=right,
        legend entries={baseline\hphantom{tec}\!\!\!\!\!\!\! BMI, selected\hphantom{ain}\!\!\!\!\!\!\! BMI}
        ]
    \addplot[ybar interval, mark=no, draw=black, area legend] plot coordinates 
    { % baseline BMI
    (14, 1) (15, 1) (16, 3) (17, 12) (18, 32) (19, 52) (20, 85) (21, 98) (22, 96) (23, 85) (24, 102) (25, 73) (26, 71) (27, 72) (28, 65) (29, 60) (30, 64) (31, 47) (32, 56) (33, 38) (34, 31) (35, 35) (36, 34) (37, 22) (38, 20) (39, 26) (40, 17) (41, 16) (42, 17) (43, 13) (44, 13) (45, 9) (46, 8) (47, 4) (48, 5) (49, 3) (50, 3) (51, 1) (52, 2) (53, 1) (54, 1)
    };
    \addplot[ybar interval, mark=no, draw=black, fill=black!50, opacity=0.6, area legend] plot coordinates 
    { % linked BMI
    (14, 0) (15, 1) (16, 0) (17, 4) (18, 15) (19, 28) (20, 73) (21, 55) (22, 54) (23, 41) (24, 10) (25, 30) (26, 22) (27, 32) (28, 27) (29, 30) (30, 24) (31, 13) (32, 16) (33, 20) (34, 5) (35, 10) (36, 29) (37, 13) (38, 6) (39, 13) (40, 14) (41, 9) (42, 3) (43, 10) (44, 6) (45, 1) (46, 1) (47, 4) (48, 5) (49, 0) (50, 1) (51, 1) (52, 0) (53, 0) (54, 0)
    };\end{axis}
    \end{tikzpicture}
    \begin{tikzpicture}
    % BRL AGE
    \begin{axis}[
        width=8.5cm,  
        height=4cm,
        ymin=-35, ymax=385,
        xmin=40, xmax=55,
        ytick={0,70,140,210,280,350},
        yticklabels={,,\hphantom{0.4},,,},
        ylabel={Frequency \textit{BRL}},
        ylabel style={yshift=-1.5pt},
        xtick style={opacity=0},
        xtick={42.5,45,47.5,50,52.5},
        xticklabels={42,45,47,50,52},
        legend style={draw=none, anchor=south west, fill=none, at={(0.6,0.55)}},
        legend plot pos=right,
        legend entries={baseline\hphantom{tec}\!\!\!\!\!\!\! Age, selected\hphantom{ain}\!\!\!\!\!\!\! Age}
        ]
    \addplot[ybar interval, mark=no, draw=black, area legend] plot coordinates 
    { % baseline AGE
    (42,330) (43,168) (44,167) (45,171) (46,173) (47,128) (48,121) (49,64) (50,43) (51,26) (52,2) (53,2)
    };
    \addplot[ybar interval, mark=no, draw=black, fill=black!50, opacity=0.6, area legend] plot coordinates 
    { % linked AGE
    (42,124) (43,43) (44,56) (45,45) (46,98) (47,51) (48,55) (49,49) (50,43) (51,8) (52,1) (53,1) 
    };\end{axis}
    \end{tikzpicture}
    \hspace{10pt}
    \begin{tikzpicture}
    % FLEXRL AGE
    \begin{axis}[
        width=8.5cm,  
        height=4cm,
        ymin=-35, ymax=385,
        xmin=40, xmax=55,
        ytick={0,70,140,210,280,350},
        yticklabels={,,\hphantom{0.4},,,},
        ylabel={Frequency \textit{FlexRL}},
        ylabel style={yshift=-1.5pt},
        xtick style={opacity=0},
        xtick={42.5,45,47.5,50,52.5},
        xticklabels={42,45,47,50,52},
        legend style={draw=none, anchor=south west, fill=none, at={(0.6,0.55)}},
        legend plot pos=right,
        legend entries={baseline\hphantom{tec}\!\!\!\!\!\!\! Age, selected\hphantom{ain}\!\!\!\!\!\!\! Age}
        ]
    \addplot[ybar interval, mark=no, draw=black, area legend] plot coordinates 
    { % baseline AGE
    (42,330) (43,168) (44,167) (45,171) (46,173) (47,128) (48,121) (49,64) (50,43) (51,26) (52,4) (53,4)
    };
    \addplot[ybar interval, mark=no, draw=black, fill=black!50, opacity=0.6, area legend] plot coordinates 
    { % linked AGE
    (42,149) (43,47) (44,58) (45,51) (46,99) (47,64) (48,53) (49,47) (50,43) (51,10) (52,1) (53,1)
    };\end{axis}
    \end{tikzpicture}
    \caption{Complementary metrics illustrating the trade-off between minimisation of linkage errors and maximisation of resemblance between linked and baseline samples, for record linkage on the SWHAN data (\textit{BRL} left, \textit{FlexRL} right). Black lines indicate means and grey bands standard deviations across 100 procedures.}
    \label{fig:tradeoff}
\end{figure}

Following these guidelines, we illustrate the selection of the cutoff $\lambda$ for our study of women's health application on \Cref{fig:tradeoff}, running the linkage procedure across a grid $\lambda \in [0.5,1)$ (Step 1) and tracking sample size on the upper x-axis. We first examine the linkage output (Step 2) via the estimated $\widehat{\text{FDP}} \in [0,1]$, computed from linkage scores \citep{fastLink_paper_2019, robachfdp}, providing an empirical order of magnitude for the attenuation bias characterised in \Cref{result_2}. As the linked sample shrinks, the estimation becomes less reliable, but remains useful for guiding selection and determining whether inference should be abandoned. We then scrutinise the linked data (Step 3) using the Maximum Mean Discrepancy, MMD $\in [0,1]$ \citep{MMDgretton12}, measuring joint distributional differences between $\mathcal{S}$ and $\mathcal{A}$ in a Gaussian RBF kernel space, providing an empirical counterpart to the sampling bias derived in \Cref{sec bias decompo}. We complement this with the Intersection over Union (IoU) of propensity score supports and covariate Standardised Mean Differences (SMD), and inspect histograms of some causally relevant variables (lower panels of \Cref{fig:tradeoff}). We additionally report the true FDP and sensitivity.

Based on these diagnostics, we select the cutoff $\lambda=0.82$, pointed out on the upper panel of \Cref{fig:tradeoff}, where linked and baseline samples begin to diverge, controlling $\widehat{\text{FDP}}$ and retaining enough observations per treatment arm (Step 4). The proportion of treated units remains stable across thresholds $\lambda$, consistent with $S \indep T | \boldsymbol{X}$. \KR{At very strict cutoffs, the IoU becomes unstable, signalling a potentially irreversible covariate shift under which all estimators become dominated by sampling bias.}

Navigating record linkage performance and representativeness of the linked sample relative to the baseline, subject to the limited number of linked observations is important for consistency to be tenable and for the effect to be recoverable. On \Cref{fig:SWHAN} and \Cref{tab:SWHAN} in the following section, we present the estimation results for the effect of physical activity on two physiological outcomes.

\section{SWAN application results} \label{sec application}

The SWAN data come from a longitudinal Study of Women's Health Across the Nation in the United States and are distributed by the Inter-university Consortium for Political and Social Research (ICPSR). We build the baseline data set $\mathcal{A}$ using the baseline (1996-1997) and visit 1 (1997-1999), and we use visit 2 (1998-2000) as the follow-up data set $\mathcal{B}$. File $\mathcal{A}$ contains 1,393 records, file $\mathcal{B}$ 1,836 and 1,194 entities are observed in both files (they form the set of true links). As an illustration, we consider the effect of initiating physical activity on two outcome variables: the hip bone mineral density (HP-BMD) and the (rescaled) dehydroepiandrosterone sulfate (DHA-S) hormone, among pre-menopausal women. In the baseline and in the true links 22\% of the units are treated.

These data are of particular interest for this work because they are openly available, suited for causal inference, fitting with the setting we tackle and they contain unique identifiers that allow to evaluate the causal record linkage procedure and discuss methodology. We work with ethnicity, age, discretised height and BMI, and a truncated portion of the unique identifier present in both data sets, as linking variables. Exact matching on all these linking variables yields an FDP of $0.3$ and a sensitivity of $0.5$. Based on the literature and on an exploratory analysis of the true links, we include baseline HP-BMD and DHA-S levels, ethnicity, BMI, age as covariates in the causal analysis \citep{sport_bonedensity_2005, sport_hormone_2005, sport_bonedensity_2015, sport_hormone_2015, sport_bonedensity_2020, sport_hormone_2023}. 

After reviewing the available software for record linkage we select \textit{BRL} and \textit{FlexRL} \citep{sadinle_bay_bipartite_RL_2017, robachflexRL} for evaluation, as they are state-of-the-art methods available in \texttt{R} representing the two main paradigms in record linkage: the Fellegi-Sunter model and the hierarchical entity resolution framework \citep{FS_RL_1969, tancredi_bay_RL_2011} and they are able to link records on this application. Following guidelines from \Cref{sec linkage selection} we run 100 procedures for both algorithms with the MMD elbow detection yielding $\lambda=0.82$, on average achieving an FDP of $0.1$ and retaining approximately $500$ linked observations---see \Cref{fig:tradeoff}.

\begin{figure}
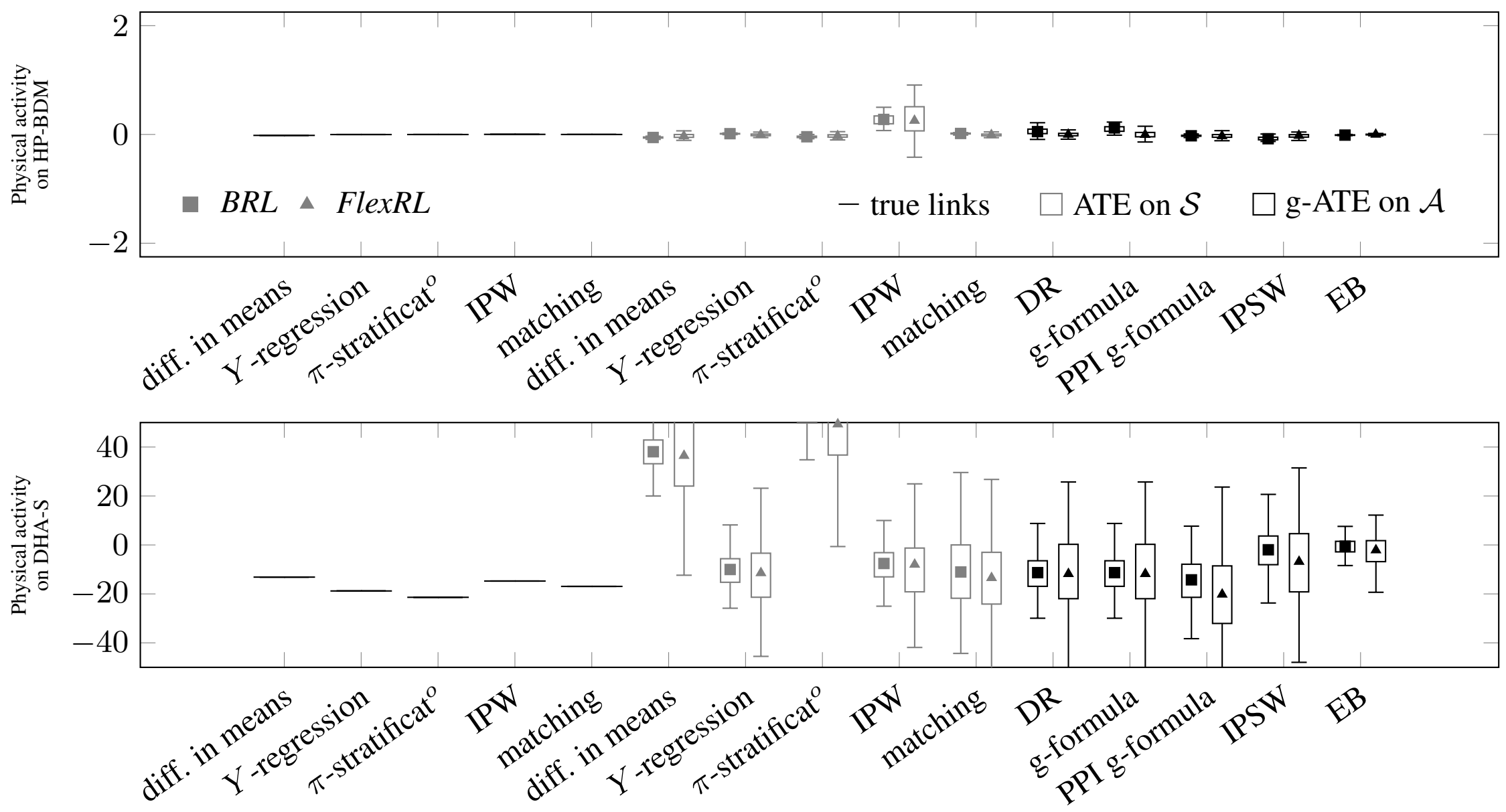

    \centering

    \input{Figures/swhan_boxplot_g_methods_2}
    
    \input{Figures/swhan_boxplot_g_methods}

\caption{Effect of physical activity on hip bone mineral density (HP-BMD) (top, no effect, average ATE on true links is $0$) and on dehydroepiandrosterone sulfate (DHA-S) hormone levels (bottom, negative effect, average ATE on true links is $-18$) over 100 procedures, after record linkage through \textit{BRL} (median indicated by a square, left point) and \textit{FlexRL} (median indicated by a triangle, right point), on the SWAN data. We compare ATE and g-ATE estimators on selected linked data $\mathcal{S}$ and on baseline data $\mathcal{A}$ respectively.}
\label{fig:SWHAN}
\end{figure}

\begin{table}
    \centering 
    % bias for the mean and standard error
    % for brl
    % for flexrl
    Physical activity on HP-BDM \hspace{50pt} \hphantom{DHA} Physical activity on DHA-S \hphantom{D} \\
    \begin{tabular}{c}
    \parbox[t]{2mm}{\rotatebox[origin=c]{90}{\!\!\!\!\!\! ATE on $\mathcal{S}$ \hphantom{......}}} \\
    \\
    \parbox[t]{2mm}{\rotatebox[origin=c]{90}{g-ATE on $\mathcal{A}$ \hphantom{..}}} \\
    \end{tabular}
    \begin{tabular}{|c|c|c|} % Y2 HP-BDM
        \hline
         estimator & \textit{BRL} & \textit{FlexRL} \\
         \hline
         diff. in means & -0.04 (0.01) & -0.01 (0.03) \\ 
         $Y$-regression & 0.01 (0.01) & -0.01 (0.02) \\        
         $\pi$-stratificat$^o$ & -0.04 (0.02) & -0.03 (0.03) \\    
         IPW & 0.27 (0.10) & 0.28 (0.27) \\     
         matching & 0.02 (0.01) & 0.01 (0.02) \\
         \hline
         DR & 0.07 (0.09) & 0.00 (0.05) \\
         g-formula & 0.11 (0.05) & 0.00 (0.08) \\
         PPI g-formula & -0.02 (0.01) & -0.03 (0.05) \\
         IPSW & -0.07 (0.03) & -0.02 (0.03) \\ % $\mathcal{S}$ \texttt{rf} 
         EB & -0.01 (0.01) & 0.00 (0.01) \\
         \hline
    \end{tabular}
    \begin{tabular}{|c|c|c|} % Y1 DHA-S Hormone
        \hline
         estimator & \textit{BRL} & \textit{FlexRL} \\
         \hline
         diff. in means        & 51.85 (9.25)     & 51.34 (20.06) \\ 
         $Y$-regression         & 9.06 (7.76)     & 7.35 (14.53) \\        
         $\pi$-stratificat$^o$ & 79.11 (11.47)     & 71.39 (20.13) \\    
         IPW                   & 7.03 (7.24)    & 5.45 (13.89) \\     
         matching              & 5.89 (16.20)     & 3.72 (18.06) \\
         \hline
         DR                    & 5.96 (8.53)  & 5.68 (17.85) \\
         g-formula             & 5.96 (8.53) & 5.68 (17.85) \\
         PPI g-formula         & 2.11 (10.44)  & -3.08 (17.63) \\
         IPSW                  & 15.36 (9.28)  & 8.84 (17.48) \\ % $\mathcal{S}$ \texttt{rf}
         EB                    & 16.51 (3.34)   & 14.42 (6.32) \\
         \hline
    \end{tabular}
    \caption{Bias mean and standard deviation over a 100 procedures of evaluating the effect of physical activity on HP-BMD (left) and on DHA-S (right) for \textit{BRL} and \textit{FlexRL} on the SWAN data. ATE and g-ATE estimators are derived from selected linked data $\mathcal{S}$ and baseline population $\mathcal{A}$ respectively.
    }
    \label{tab:SWHAN}
\end{table}

\Cref{fig:SWHAN} presents the distributions of the estimated effect of physical activity on both outcomes, HP-BMD and DHA-S, for both record linkage algorithms, \textit{BRL} and \textit{FlexRL}. For each outcome, we compare standard ATE estimators on the selected linked data $\mathcal{S}$ (cited in \Cref{sec bias decompo}) with generalised treatment effect estimators, g-ATE, on the baseline population $\mathcal{A}$ (specified in \Cref{sec estimation}) to ATE in the true links; corresponding average biases are reported in \Cref{tab:SWHAN}. Across linkage algorithms, the biases of all estimators are of similar magnitude.

There is no effect of physical activity on HP-BMD. All estimators tally around zero; correctly and incorrectly linked pairs contribute similarly in absence of signal. In contrast, there is a negative effect of physical activity on DHA-S. On \Cref{tab:SWHAN} we see that the `matching' ATE estimator and `PPI g-formula' g-ATE estimator yield the most consistent estimates for both record linkage algorithms. We observe that \textit{FlexRL} is more noisy and less conservative than \textit{BRL}, which is expected \citep{robachflexRL, robachfdp}.

Interestingly, weighting g-methods (IPSW, EB) are not particularly sensitive to positivity and overlap violations in this application. All estimators seem robust to the proportion of false discoveries in the linked data, suggesting some tolerance to deviations from \Cref{ass fdp}. However, we noticed that all estimators become dominated by the sampling bias with an overly strict linkage decision rule. The true effect average is around $-18$ which we approach with $\lambda=0.82$. With $\lambda=0.95$ estimators estimate an effect of $+15$; we also see that the MMD deviates from the null with the increasing stringency of the decision rule on \Cref{fig:tradeoff} and that the IoU of propensity score support shrinks.

Overall, this application illustrates the practical difficulties of causal record linkage: linkage quality and linked observations after selection are limited. We complement the application with a simulation study demonstrating the practical utility of our methodology in the following section.

\section{Simulations} \label{sec simulations}

To better understand the estimation behaviour in the causal record linkage framework, we perform a simulation study. Files $\mathcal{A}$ and $\mathcal{B}$ gather respectively 500 and 700 patient records, including 400 true links created by randomly duplicating records between sources. We generate five independent variables $\boldsymbol{X}$ from multinomial distributions. We define the propensity score as $\pi(\boldsymbol{X}) = \mathbb P (T=1|\boldsymbol{X}) = \Lambda(\gamma X_1 + \gamma X_2 + \gamma X_4)$, where $\Lambda(\cdot)$ is the logistic distribution function. We define the outcome as $Y(t) = \alpha t X_1 + \beta X_2 + \varepsilon$ incorporating effect heterogeneity through an interaction with $X_1$. While $\boldsymbol{C} = (X_1,X_2,X_4)$ are involved in the causal framework, the record linkage model uses $\boldsymbol{Z} = (X_1,X_2,X_3,X_5)$ to link the data sources. Variables $\boldsymbol{X}$ have a decreasing distribution so that higher values are rarer; we do so to illustrate that, although noise in the linked data is expected to shrink the effect towards zero, sampling bias can inflate it if the profiles record linkage preferentially selects are associated with a larger causal effect. Conversely, when selected profiles are associated with a smaller causal effect, sampling bias reinforces the attenuation bias; this is what we observe empirically for the null effect of physical activity on HP-BMD in \Cref{sec application}.

We analyse a scenario with good linkage quality (FDP after selection below $0.05$); we obtained similar results with more difficult scenarios (FDP after selection below $0.2$). Linking variables have a relatively high discriminative strength with 7 unique values each, and we do not induce any registration error. Exact matching on all linking variables yields an FDP of $0.5$ and a sensitivity of $1$. This simulation scenario results in a default FDP of $0.15$ for both methods \textit{BRL} and \textit{FlexRL} linking 290 and 330 records respectively, and an optimised FDP of $0.03$ for both methods \textit{BRL} and \textit{FlexRL} linking 110 records and 130 records respectively. The \textit{BRL} method is known to be more conservative \citep{robachflexRL, robachfdp}, and that is indeed what we observe.

\begin{figure}
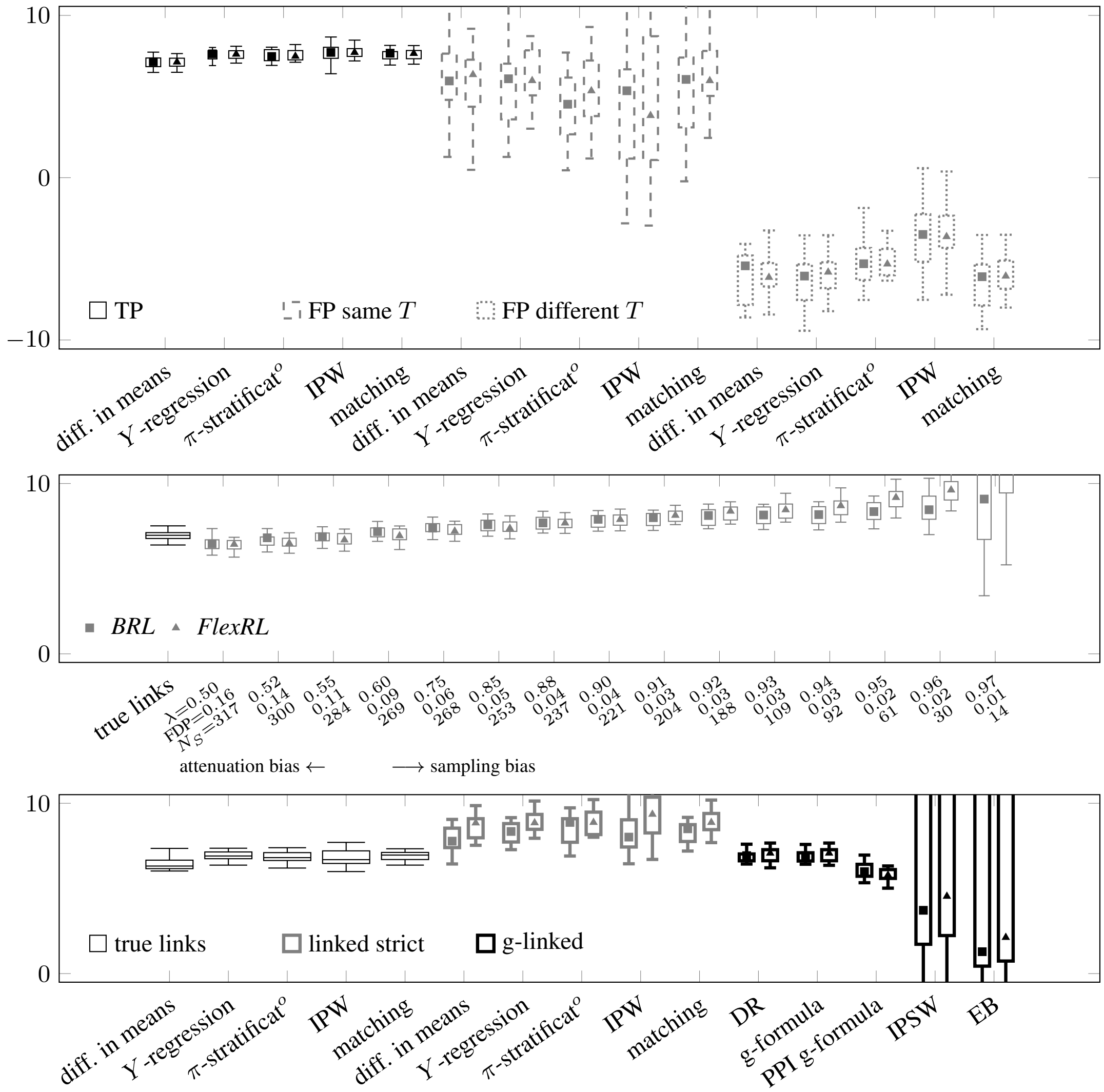

    \centering

    % decomposition tp fp BRL, FlexRL
    \input{Figures/simu_decompo_tp_fp}
    
    % evolution of the g-formula with the stringency of the linkage
    \input{Figures/simu_ymodelling_evol}
    
    % g-methods
    \input{Figures/simu_g_methods}
    
\caption{ATE of the simulation study over 100 procedures, after record linkage through \textit{BRL} (square, left point) and \textit{FlexRL} (triangle, right point). On the top, estimators evaluated on TP and FP (of the same or of opposite treatment group) for $\lambda=0.5$. On the middle, evolution of outcome regression estimator ($Y$-regression) with linkage stringency (assessed by $\lambda$, the resulting FDP and the number of linked observations $N_S$). On the bottom, ATE estimators evaluated on true links and selected linked data $\mathcal{S}$ with $\lambda=0.95$ and, g-ATE estimators evaluated on baseline data $\mathcal{A}$.}
\label{fig:simu}
\end{figure}

\begin{table}
    \centering 
    % bias for the mean and standard error
    % for brl
    % for flexrl
    \begin{tabular}{c}
    \parbox[t]{2mm}{\rotatebox[origin=c]{90}{\!\!\!\!\!\! ATE on $\mathcal{S}$ \hphantom{......}}} \\
    \\
    \parbox[t]{2mm}{\rotatebox[origin=c]{90}{g-ATE on $\mathcal{A}$ \hphantom{..}}} \\
    \end{tabular}
    \begin{tabular}{|c|c|c|}
        \hline
         estimator & \textit{BRL} & \textit{FlexRL} \\
         \hline
         diff. in means & 1.27 (0.63) & 2.00 (0.60) \\ 
         $Y$-regression & 1.26 (0.52) & 1.81 (0.51) \\        
         $\pi$-stratificat$^o$ & 1.31 (0.73) & 1.84 (0.60) \\    
         IPW & 1.42 (1.07) & 2.19 (1.21) \\     
         matching & 1.28 (0.51) & 1.79 (0.54) \\ 
         \hline
         DR & 0.10 (0.29) & 0.18 (0.37) \\ 
         g-formula & 0.10 (0.29) & 0.19 (0.35) \\
         PPI g-formula & -0.63 (0.47) & -0.89 (0.53) \\
         IPSW & --- & --- \\ % $\mathcal{S}$ \texttt{rf}
         EB & --- & --- \\ 
         \hline
    \end{tabular}
    \caption{Bias mean and standard deviation over a 100 procedures of the simulation study for \textit{BRL} and \textit{FlexRL}. ATE and g-ATE estimators are obtained on selected linked data and baseline population respectively.}
    \label{tab:simu}
\end{table}

The simulation study illustrates the discussions from \Cref{sec identifiability} and \Cref{sec bias decompo}. Using the default linkage threshold $\lambda=0.5$, ATE estimators exhibit attenuation bias from falsely linked data due to covariate discrepancies and adversarial contributions---see the upper \Cref{fig:simu}. Falsely linked data yield larger variance than correctly linked data. 

The trade-off between attenuation bias and sampling bias is illustrated for the outcome regression estimator in the middle panel of \Cref{fig:simu}; we observed the same behaviour for all estimators. These results illustrate that deviations from the main effect or increasing variance may not reflect inconsistencies in the linked data but rather may arise from sampling bias; highlighting limitations of approaches aiming at minimising the estimated effect variance \citep{wortman_RL_causality_2018}. In the middle panel of \Cref{fig:simu}, sampling bias induces larger variance and deviations from the main effect despite an FDP lower than $0.03$.

Optimising MMD and controlling the estimated FDP yields $\lambda=0.95$ for \textit{BRL} and \textit{FlexRL}. We report in \Cref{tab:simu} the average biases of ATE estimators on selected linked data $\mathcal{S}$ and g-ATE estimators on baseline population $\mathcal{A}$. While the estimation from rigorously selected linked data (middle of the lower panel on \Cref{fig:simu}) is biased, g-methods relying on outcome modelling (right of the lower panel on \Cref{fig:simu}) recover the true effect. Weighting methods though are affected by the lack of positivity and overlap between the covariate distributions of $\mathcal{S}$ and $\mathcal{A}$ and are not reliable. 

\section{Conclusion and Discussion} \label{sec discussion} 

This paper frames causal inference on linked data as a selection problem that primarily threatens consistency. Under a negligible false discovery proportion or a valid linking variable adjustment set, the selection backdoor criterion ensures effect identifiability and transportability to the target. We further highlight the balance between linkage reliability and sample representativeness. To address the bias we advocate for the use of g-methods, whose assumptions offer different strategies for causal inference and can be challenged using generalised additive models or sensitivity analysis. We provide a principled framework for selecting linked data and estimating causal effects while accounting for the trade-offs inherent in record linkage and generalisability.

When record linkage is imperfect, it remains capable of matching similar profiles and allows the treatment effect to be recovered with g-methods, provided that the linking variables overlap sufficiently with the confounders. Reliable causal estimation thus remains possible despite linkage errors to some extent. This connects record linkage to matching, though unlike matching methods with an adaptable number of matches, record linkage as we studied it does not benefit from building clusters of comparable entities, as it aims to recover the one-to-one assignment between entities across two data files. Moreover, record linkage performance improves with the dimensionality of linking variables, whereas matching discrepancy worsens with the dimensionality of covariates. Bridging these two paradigms further might be a promising direction for future work on causal record linkage.

The selection problem we investigated is not unique to causal inference and would appear for any statistical inference on linked data. More broadly, the causal record linkage framework is an extreme instance of the classical problem of learning under a non-ignorable selection mechanism that shifts the observed (`training') distribution away from the target population. Our case is particularly severe because record linkage selection targets a narrow, atypical subpopulation; were the selected population representative of the majority of the baseline observations, the selection mechanism would remain non-ignorable, but the covariate distributions from which we learn and to which we generalise would overlap far more.

Uncertainty is not addressed in this article. Horvitz-Thompson or robust variance estimators cannot account for the uncertainty arising from the entire process (data sampling, linkage, propensity score and treatment effect estimation), and variance estimators derived under simple random sampling are inappropriate because linkage is informed by covariates. Ideally, we would propagate uncertainty via bootstrapping. However, as with matching methods \citep{NoBootstrap_2008}, the bootstrap is not valid for causal record linkage because it cannot preserve the distributions of links and non-links while maintaining independence across samples. Moreover, it cannot ensure adequate representation of treated and control units given the limited number of linked observations available in each sample. Therefore, we only report point estimates in the applications and leave the quantification of uncertainty throughout the causal record linkage process for future research. Notably, the argument for what to do in practice may change when uncertainty quantification is desired, trading bias for variance reduction.

Explainability and diagnostic tools are difficult to develop because causal relationships are only observed for linked records, and only partial information is available from non-linked records (covariates in the baseline and outcomes in the follow-up). We therefore proposed diagnostic measures to help scientists understand how atypical the linked population is, report on the precariousness of the estimated treatment effect across the space of confounders, and clarify whether uncertainty stems from data scarcity or reliance on modelling. With this essay on causal record linkage, we hope to have provided a set of concepts that can anchor future methodologies for causal inference on linked data.

\begin{appendix}

\section{Appendix} \label{sec appendix}

\subsection{Standard average treatment effect estimators} \label{appendix ate}

\begin{align*}
    \text{difference in means: } & \frac{1}{N_{S1}} \sum_{i,j} S_{i,j} \, T_i \, Y_j - \frac{1}{N_{S0}} \sum_{i,j} S_{i,j} \, (1-T_i) \, Y_j ,\\
    \text{regression estimator: } & \frac{1}{N_{S}} \sum_{i,j} S_{i,j} \, \big( \widehat{\mu}_1(\boldsymbol{X}_i) - \widehat{\mu}_0(\boldsymbol{X}_i) \big), \text{ where } \widehat{\mu}_t(\boldsymbol{X}_i) \text{ estimates } \mathbb E [Y_j|T_i=t,\boldsymbol{X}_i],\\
    \text{propensity score stratification: } & \sum_{\text{stratum } s \in \mathcal{S}} \Big( \frac{1}{N_{S1}^s} \sum_{\{i,j\} \in s} T_i \, Y_j - \frac{1}{N_{S0}^s} \sum_{\{i,j\} \in s} (1-T_i) \, Y_j \Big) \, \frac{N_{S}^s}{N_{S}},\\
    \text{inverse probability weighting: } & \frac{1}{N_{S}} \sum_{i,j} S_{i,j} \, \Big( \frac{T_i \, Y_j}{\widehat{\pi}(\boldsymbol{X}_i)} - \frac{(1-T_i) \, Y_j}{(1-\widehat{\pi}(\boldsymbol{X}_i))} \Big),\\
    \text{matching: } & \frac{1}{N_{S}} \sum_{i,j} (2 T_i - 1) (Y_j - \frac{1}{M} \sum_{i^\prime \in \mathcal{J}_M(i)} Y_{j^\prime}) S_{i,j}.
\end{align*}

\subsection{Proof of Theorems} \label{appendix proof}

We adopt the shorthand $S \Delta$ for TP and $S \nabla$ for FP (with $S \nabla \mathcal{A}, S \nabla \mathcal{B}$ specifying conditioning on $\boldsymbol{X}_i$ or $\boldsymbol{X}_j$, and $S \nabla t t^\prime$ disambiguating falsely linked pairs with $T_i=t$ and $T_j=t^\prime$; no such distinction is needed for TP). Additionally, potential outcomes in $\mathcal{S}$ might refer to those of records from $\mathcal{A}$ or from $\mathcal{B}$, we sometimes index these quantities with $S_{\mathcal{A}}$ or $S_{\mathcal{B}}$. We refer the reader to \Cref{sec population} for other notations.

\paragraph{Objectives} We consider the causal record linkage system presented in \Cref{sec population} and represented in \Cref{fig:SelectionDiagram}. Our estimand, the target of inference, is the Average Treatment Effect (ATE) in the baseline population (\Cref{estimand_tau_A} in the article) $$\tau_{\mathcal{A}} = \mathbb E_{\mathcal{A}} [Y(1) - Y(0)] = {\mu_1}_{\mathcal{A}} - {\mu_0}_{\mathcal{A}}.$$ Let $\widehat{\tau}_{\mathcal{S}} = \frac{1}{N_{S1}} \sum_{i,j} S_{i,j} \, T_i \, Y_j - \frac{1}{N_{S0}} \sum_{i,j} S_{i,j} \, (1-T_i) \, Y_j$ denote the difference in means estimator (\Cref{estimator_hat_tau_S} in the article), computed on the selected linked data. The bias of the causal record linkage procedure is thus: $$\mathbb E [\widehat{\tau}_{\mathcal{S}}] - \tau_{\mathcal{A}} = \mathbb E [ \widehat{\zeta}_{N_{S}} ] + \mathbb E [ \widehat{\xi}_{N_{S}} ].$$ It decomposes into the sampling bias $\widehat{\zeta}_{N_{S}}$ and the attenuation bias $\widehat{\xi}_{N_{S}}$. The sampling bias is due to the record linkage induced selection mechanism and results from the missed links while the attenuation bias is due to the falsely linked pairs, which contaminates the signal with opposite contributions owing to the mismatch in treatment groups of some falsely linked pairs and, with effect dilution owing to the discrepancies between potential outcomes of all falsely linked pairs. We remind the reader of the biases definitions given in \Cref{sec bias decompo}: \begin{align*}
    \widehat{\zeta}_{N_{S}} = & \frac{1}{N_{S1}} \sum_{i,j} \big\{ S_{i,j} \, T_i \, \mu_1(\boldsymbol{X}_i) \big\} - \frac{1}{N_{S0}} \sum_{i,j} \big\{ S_{i,j} \, (1-T_i) \, \mu_0(\boldsymbol{X}_i) \big\} - \tau_{\mathcal{A}},\\
    \widehat{\xi}_{N_{S}} = & \frac{1}{N_{S1}} \sum_{i,j} \big\{ S_{i,j} \,T_i \, (Y_j - \mu_1(\boldsymbol{X}_j)) \big\} - \frac{1}{N_{S0}} \sum_{i,j} \big\{ S_{i,j} \, (1-T_i) \, (Y_j - \mu_0(\boldsymbol{X}_j)) \big\} \\
    & + \frac{1}{N_{S0}} \sum_{i,j} \big\{ S_{i,j} \, (1-T_i) \, (\mu_0(\boldsymbol{X}_i)-\mu_0(\boldsymbol{X}_j)) \big\}  - \frac{1}{N_{S1}} \sum_{i,j} \big\{ S_{i,j} \, T_i \, (\mu_1(\boldsymbol{X}_i)-\mu_1(\boldsymbol{X}_j))\big\}.
\end{align*} Note that $\mu_{t}(\boldsymbol{X}_i)$ does not depend on the actual treatment the unit received under conditional exchangeability (see \Cref{sec exchangeability}, which holds in the selected data given $\boldsymbol{X}_i$ for both correctly and incorrectly linked pairs). The same property applies to $\mu_{t}(\boldsymbol{X}_j)$ given $\boldsymbol{X}_j$. However in practice, we cannot condition on $\boldsymbol{X}_j$ which is unobserved unless either \Cref{ass fdp} or the point of \Cref{rmk link var valid} holds.

\subsubsection{The sampling bias is the difference in ATE between data set $\mathcal A$ and its subset $\mathcal S$} 
\label{sec sampling bias}

Under assumption \ref{ass treated control}, \begin{align*}
    \mathbb E [ \widehat{\zeta}_{N_S} ] = & \mathbb E \Big[ \frac{1}{N_{S1}} \sum_{i,j} \big\{ S_{i,j} \, T_i \, \mu_1(\boldsymbol{X}_i) \big\} - \frac{1}{N_{S0}} \sum_{i,j} \big\{ S_{i,j} \, (1-T_i) \, \mu_0(\boldsymbol{X}_i) \big\} - \tau_{\mathcal{A}} \Big]\\
    = & \underset{\substack{\text{what would have happened to} \\ \text{selected treated units from $\mathcal{A}$}}}{\underbrace{\mathbb E \Big[ \frac{1}{N_{S1}} \sum_{i,j} S_{i,j} \, T_i \, \mu_1(\boldsymbol{X}_i) \Big]}} - \underset{\substack{\text{what would have happened to} \\ \text{selected untreated units from $\mathcal{A}$}}}{\underbrace{\mathbb E \Big[ \frac{1}{N_{S0}} \sum_{i,j} S_{i,j} \, (1-T_i) \, \mu_0(\boldsymbol{X}_i) \Big]}} - \tau_{\mathcal{A}} \\
    = & {\mu_1}_{\mathcal{S}_{\mathcal{A}}} - {\mu_0}_{\mathcal{S}_{\mathcal{A}}} - ({\mu_1}_{\mathcal{A}} - {\mu_0}_{\mathcal{A}}) \\
    = & \tau_{\mathcal{S}_{\mathcal{A}}} - \tau_{\mathcal{A}}.
\end{align*}

\subsubsection{The attenuation bias reflects the impact of falsely linked data on the causal inference} 
\label{sec attenuation bias}

Falsely linked data impact causal inference, with adversarial contributions from linking records that pertain to different treatment groups and, with discrepancies from linking records that differ in covariates: \begin{align}
    \widehat{\xi}_{N_{S}} = & \frac{1}{N_{S1}} \sum_{i,j} \big\{ S_{i,j} \, T_i \, (Y_j - \mu_1(\boldsymbol{X}_j)) \big\} - \frac{1}{N_{S0}} \sum_{i,j} \big\{ S_{i,j} \, (1-T_i) \, (Y_j - \mu_0(\boldsymbol{X}_j)) \big\} \label{opposit contrib} \\
    & + \frac{1}{N_{S0}} \sum_{i,j} \big\{ S_{i,j} \, (1-T_i) \, (\mu_0(\boldsymbol{X}_i)-\mu_0(\boldsymbol{X}_j)) \big\} - \frac{1}{N_{S1}} \sum_{i,j} \big\{ S_{i,j} \, T_i \, (\mu_1(\boldsymbol{X}_i)-\mu_1(\boldsymbol{X}_j)) \big\}  \label{attenuation} \\
    = & \frac{1}{N_{S1}} \sum_{i,j} \big\{ S_{i,j} \, \Delta_{i,j} \, T_i \, ( Y_j - \mu_1(\boldsymbol{X}_j) ) \big\} \tag{9}\\
    & - \frac{1}{N_{S0}} \sum_{i,j} \big\{ S_{i,j} \, \Delta_{i,j} \, (1-T_i) \, ( Y_j - \mu_0(\boldsymbol{X}_j) ) \big\} \tag{9} \label{TP opposit contrib} \\
    & + \frac{1}{N_{S1}} \sum_{i,j} \big\{ S_{i,j} \, (1-\Delta_{i,j}) \, T_i \, (Y_j-\mu_1(\boldsymbol{X}_j)) \big\} \tag{10} \\
    & - \frac{1}{N_{S0}} \sum_{i,j} \big\{ S_{i,j} \, (1-\Delta_{i,j}) \, (1-T_i) \, (Y_j-\mu_0(\boldsymbol{X}_j)) \big\} \tag{10} \label{FP opposit contrib} \\
    & + \frac{1}{N_{S0}} \sum_{i,j} \big\{ S_{i,j} \, \Delta_{i,j} \, (1-T_i) \, (\mu_0(\boldsymbol{X}_i)-\mu_0(\boldsymbol{X}_j)) \big\} \tag{11} \\
    & - \frac{1}{N_{S1}} \sum_{i,j} \big\{ S_{i,j} \, \Delta_{i,j} \, T_i \, (\mu_1(\boldsymbol{X}_i)-\mu_1(\boldsymbol{X}_j)) \big\} \tag{11} \label{TP attenuation} \\
    & + \frac{1}{N_{S0}} \sum_{i,j} \big\{ S_{i,j} \, (1-\Delta_{i,j}) \, (1-T_i) \, (\mu_0(\boldsymbol{X}_i)-\mu_0(\boldsymbol{X}_j)) \big\} \tag{12} \\
    & - \frac{1}{N_{S1}} \sum_{i,j} \big\{ S_{i,j} \, (1-\Delta_{i,j}) \, T_i \, (\mu_1(\boldsymbol{X}_i)-\mu_1(\boldsymbol{X}_j)) \big\} \tag{12} \label{FP attenuation}.
\end{align} Components (\ref{opposit contrib}) and (\ref{attenuation}) separate into TP/FP in (\ref{TP opposit contrib})/(\ref{FP opposit contrib}) and (\ref{TP attenuation})/(\ref{FP attenuation}) respectively. By consistency in the source population (\Cref{key ass hold}), there is no difference between observed and potential outcomes in TP, and component (\ref{TP opposit contrib}) vanishes. Likewise, since TP refer to the same individuals, their potential outcomes in $\mathcal{A}$ and in $\mathcal{B}$ coincide, and component (\ref{TP attenuation}) vanishes. In contrast, FP between records from different treatment groups induce opposite contributions (\ref{FP opposit contrib}) and, all FP induce bias through the discrepancy between their baseline and follow-up potential outcomes (\ref{FP attenuation}).

\setcounter{equation}{12}

From the difference between observed and potential outcomes of falsely linked records from different treatment groups arises adversarial contributions; for each treatment group from lines (\ref{FP opposit contrib}): \begin{align*}
    & \mathbb E \Big[ \frac{1}{{N_{S}}_t} \sum_{i,j} S_{i,j} \, (1-\Delta_{i,j}) \, \ind_{\{ T_i = t\}} \, (Y_j-\mu_t(\boldsymbol{X}_j)) \Big] \\
    = & \mathbb E \Big[ \frac{1}{{N_{S}}_t} \sum_{i,j} S_{i,j} \, (1-\Delta_{i,j}) \, \ind_{\{ T_i = t\}} \, \ind_{\{ T_j = t\}} \, (Y_j-\mu_t(\boldsymbol{X}_j)) \Big] \\
    & + \mathbb E \Big[ \frac{1}{{N_{S}}_t} \sum_{i,j} S_{i,j} \, (1-\Delta_{i,j}) \, \ind_{\{ T_i = t\}} \, \ind_{\{ T_j = 1-t\}} \, (Y_j-\mu_t(\boldsymbol{X}_j)) \Big] \\
    = & \mathbb E \Big[ \frac{1}{{N_{S}}_t} \sum_{i,j} S_{i,j} \, (1-\Delta_{i,j}) \, \ind_{\{ T_i = t\}} \, \ind_{\{ T_j = t\}} \, \big\{ \mathbb E [Y_j|T_j=t,\boldsymbol{X}_j]-\mu_t(\boldsymbol{X}_j) \big\} \Big] \\
    & + \mathbb E \Big[ \frac{1}{{N_{S}}_t} \sum_{i,j} S_{i,j} \, (1-\Delta_{i,j}) \, \ind_{\{ T_i = t\}} \, \ind_{\{ T_j = 1-t\}} \, \big\{ \mathbb E [Y_j|T_j=1-t,\boldsymbol{X}_j]-\mu_t(\boldsymbol{X}_j)) \big\} \Big] \\
    = & \mathbb E \Big[ \frac{1}{{N_{S}}_t} \sum_{i,j} S_{i,j} \, (1-\Delta_{i,j}) \, \ind_{\{ T_i = t\}} \, \ind_{\{ T_j = t\}} \, \big\{ \mathbb E [Y_j(t)|\boldsymbol{X}_j]-\mu_t(\boldsymbol{X}_j) \big\} \Big] \\
    & + \mathbb E \Big[ \frac{1}{{N_{S}}_t} \sum_{i,j} S_{i,j} \, (1-\Delta_{i,j}) \, \ind_{\{ T_i = t\}} \, \ind_{\{ T_j = 1-t\}} \, \big\{ \mathbb E [Y_j(1-t)|\boldsymbol{X}_j]-\mu_t(\boldsymbol{X}_j) \big\} \Big] \\
    = & \mathbb E \Big[ \frac{1}{{N_{S}}_t} \sum_{i,j} S_{i,j} \, (1-\Delta_{i,j}) \, \ind_{\{ T_i = t\}} \, \ind_{\{ T_j = t\}} \, (\mu_t(\boldsymbol{X}_j)-\mu_t(\boldsymbol{X}_j)) \Big] \\
    & + \mathbb E \Big[ \frac{1}{{N_{S}}_t} \sum_{i,j} S_{i,j} \, (1-\Delta_{i,j}) \, \ind_{\{ T_i = t\}} \, \ind_{\{ T_j = 1-t\}} \, (\mu_{1-t}(\boldsymbol{X}_j)-\mu_t(\boldsymbol{X}_j)) \Big] \\
    = & \mathbb E \Big[ \frac{1}{{N_{S}}_t} \sum_{i,j} S_{i,j} \, (1-\Delta_{i,j}) \, \ind_{\{ T_i = t\}} \, \ind_{\{ T_j = 1-t\}} \, (\mu_{1-t}(\boldsymbol{X}_j)-\mu_t(\boldsymbol{X}_j)) \Big] \\
    = & \mathbb E \Big[ \frac{N_{S \nabla t, 1-t}}{{N_{S}}_t} \frac{1}{{N_{S \nabla t, 1-t}}} \sum_{i,j} S_{i,j} \, (1-\Delta_{i,j}) \, \ind_{\{ T_i = t\}} \, \ind_{\{ T_j = 1-t\}} \, ( \mu_{1-t}(\boldsymbol{X}_j)-\mu_t(\boldsymbol{X}_j) ) \Big] \\
    = & \mathbb E \Big[ \frac{N_{S \nabla t, 1-t}}{{N_{S}}_t} \Big] ( {\mu_{1-t}}_{S\nabla_{\mathcal{B}}}-{\mu_t}_{S\nabla_{\mathcal{B}}} ).
\end{align*} We use the consistency of the ATE estimator (\Cref{ass consisctency ate}) to replace its empirical average by its expectation, and the consistency in the source population (\Cref{key ass hold}) to replace observed outcomes by potential outcomes. Then, component (\ref{FP opposit contrib}) of $\widehat{\xi}_{{N_{S}}}$ develops into: \begin{align*}
    & \mathbb E \Big[ \frac{1}{N_{S1}} \sum_{i,j} S_{i,j} \, (1-\Delta_{i,j}) \, \ind_{\{ T_i = 1\}} \, (Y_j-\mu_1(\boldsymbol{X}_j)) \Big] - \mathbb E \Big[ \frac{1}{N_{S0}} \sum_{i,j} S_{i,j} \, (1-\Delta_{i,j}) \, \ind_{\{ T_i = 0\}} \, (Y_j-\mu_0(\boldsymbol{X}_j)) \Big] \\
    = & \mathbb E \Big[ \frac{N_{S \nabla 1, 0}}{N_{S1}} \Big] ( {\mu_{0}}_{S\nabla_{\mathcal{B}}}-{\mu_1}_{S\nabla_{\mathcal{B}}} ) - \mathbb E \Big[ \frac{N_{S \nabla 0, 1}}{N_{S0}} \Big] ( {\mu_{1}}_{S\nabla_{\mathcal{B}}}-{\mu_0}_{S\nabla_{\mathcal{B}}} ) \\
    = & \mathbb E \Big[ \frac{N_{S \nabla 1, 0}}{N_{S1}} + \frac{N_{S \nabla 0, 1}}{N_{S0}} \Big] \: ({\mu_0}_{S\nabla_{\mathcal{B}}} - {\mu_1}_{S\nabla_{\mathcal{B}}} \vphantom{\frac{N_{S \nabla 1, 0}}{N_{S1}}}).
\end{align*} We observe the opposite contributions, reflecting the opposite of the treatment effect. It arises when falsely linked records belong to different treatment groups. This bias involves the potential outcomes of records from the follow-up data that are falsely linked, defined over the $\boldsymbol{X}$ from $\mathcal{B}$ which we do not have access to.

From the information discrepancy in falsely linked records arises a difference between potential outcomes from $\mathcal{A}$ and from $\mathcal{B}$; for each treatment group from lines (\ref{FP attenuation}): \begin{align*}
    & \mathbb E \Big[ \frac{1}{{N_{S}}_t} \sum_{i,j} S_{i,j} \, (1-\Delta_{i,j}) \, \ind_{\{ T_i = t\}} \, (\mu_t(\boldsymbol{X}_i)-\mu_t(\boldsymbol{X}_j)) \Big] \\
    = & \mathbb E \Big[ \frac{1}{{N_{S}}_t} \sum_{i,j} S_{i,j} \, (1-\Delta_{i,j}) \, \ind_{\{ T_i = t\}} \, \ind_{\{ T_j = t\}} \, (\mu_t(\boldsymbol{X}_i)-\mu_t(\boldsymbol{X}_j)) \Big] \\
    & + \mathbb E \Big[ \frac{1}{{N_{S}}_t} \sum_{i,j} S_{i,j} \, (1-\Delta_{i,j}) \, \ind_{\{ T_i = t\}} \, \ind_{\{ T_j = 1-t\}} \, (\mu_t(\boldsymbol{X}_i)-\mu_t(\boldsymbol{X}_j)) \Big] \\
    = & \mathbb E \Big[ \frac{N_{S \nabla t, t}}{{N_{S}}_t} \frac{1}{{N_{S \nabla t, t}}} \sum_{i,j} S_{i,j} \, (1-\Delta_{i,j}) \, \ind_{\{ T_i = t\}} \, \ind_{\{ T_j = t\}} \, (\mu_t(\boldsymbol{X}_i)-\mu_t(\boldsymbol{X}_j)) \Big] \\
    & + \mathbb E \Big[ \frac{N_{S \nabla t, 1-t}}{{N_{S}}_t} \frac{1}{{N_{S \nabla t, 1-t}}} \sum_{i,j} S_{i,j} \, (1-\Delta_{i,j}) \, \ind_{\{ T_i = t\}} \, \ind_{\{ T_j = 1-t\}} \, (\mu_t(\boldsymbol{X}_i)-\mu_t(\boldsymbol{X}_j)) \Big] \\
    = & \mathbb E \Big[ \frac{N_{S \nabla t, t}}{{N_{S}}_t} \Big] ({\mu_t}_{S\nabla_{\mathcal{A}}}-{\mu_t}_{S\nabla_{\mathcal{B}}})  \\
    & + \mathbb E \Big[ \frac{N_{S \nabla t, 1-t}}{{N_{S}}_t} \Big] ({\mu_t}_{S\nabla_{\mathcal{A}}}-{\mu_t}_{S\nabla_{\mathcal{B}}}) \\
    = & \mathbb E \Big[ \frac{N_{S \nabla t, t} + N_{S \nabla t, 1-t}}{{N_{S}}_t} \Big] ({\mu_t}_{S\nabla_{\mathcal{A}}}-{\mu_t}_{S\nabla_{\mathcal{B}}}) \\
    = & \mathbb E \big[ \text{FDP}_t \big] ({\mu_t}_{S\nabla_{\mathcal{A}}}-{\mu_t}_{S\nabla_{\mathcal{B}}}).
\end{align*} Drawing on results from the matching literature, under \Cref{rmk link var valid} and assumptions \ref{ass reg pot outcome} and \ref{ass support cov}, this discrepancy is stochastically bounded: $({\mu_t}_{S\nabla_{\mathcal{A}}}-{\mu_t}_{S\nabla_{\mathcal{B}}}) = \mathcal{O}(\frac{1}{{N_{\!\mathcal{S}}}^{\!1\!/\!k}})$. \begin{assumption} \label{ass reg pot outcome} The potential outcomes satisfy standard regularity conditions, ensuring smoothness of $\mu_t(\boldsymbol{X})$. \end{assumption} \begin{assumption} \label{ass support cov} The covariates $\boldsymbol{X} \in \mathbb R^k$ have compact, convex support, and density bounded away from zero. \end{assumption} The stochastic order of this bias is derived from the proof of theorem 1 in \citep{AandIMatching2006}.

Then, component (\ref{FP attenuation}) of $\widehat{\xi}_{{N_{S}}}$ develops into: \begin{align*}
    & \mathbb E \Big[ \frac{1}{N_{S0}} \sum_{i,j} S_{i,j} \, (1-\Delta_{i,j}) \, (1-T_i) \, (\mu_0(\boldsymbol{X}_i)-\mu_0(\boldsymbol{X}_j)) \big] \\
    & - \mathbb E \Big[ \frac{1}{N_{S1}} \sum_{i,j} S_{i,j} \, (1-\Delta_{i,j}) \, T_i \, (\mu_1(\boldsymbol{X}_i)-\mu_1(\boldsymbol{X}_j)) \big] \\
    = & \mathbb E \big[ \text{FDP}_0 \big] ({\mu_0}_{S\nabla_{\mathcal{A}}}-{\mu_0}_{S\nabla_{\mathcal{B}}}) - \mathbb E \big[ \text{FDP}_1 \big] ({\mu_1}_{S\nabla_{\mathcal{A}}}-{\mu_1}_{S\nabla_{\mathcal{B}}}).
\end{align*} This result explains the effect dilution due to incorrect information being used. It results from the discrepancy between potential outcomes from $\mathcal{A}$ and from $\mathcal{B}$ in any falsely linked pairs. 

To summarise the Appendix \Cref{sec sampling bias} and \Cref{sec attenuation bias}, we studied the bias: $$\mathbb E [\widehat{\tau}_{\mathcal{S}}] - \tau_{\mathcal{A}} = \mathbb E [ \widehat{\zeta}_{{N_{S}}} ] + \mathbb E [ \widehat{\xi}_{{N_{S}}} ],$$ and defined the sampling bias due to missed links: $$\mathbb E [ \widehat{\zeta}_{{N_{S}}} ] = {\mu_1}_{\mathcal{S}_{\mathcal{A}}} - {\mu_0}_{\mathcal{S}_{\mathcal{A}}} - ({\mu_1}_{\mathcal{A}} - {\mu_0}_{\mathcal{A}}) = \underset{\text{sampling bias}}{\underbrace{\tau_{\mathcal{S}_{\mathcal{A}}} - \tau_\mathcal{A}}},$$ as well as the attenuation bias due to falsely linked data: $$\mathbb E [ \widehat{\xi}_{{N_{S}}} ] = \mathbb E \Big[ \frac{N_{S \nabla 1, 0}}{N_{S1}} + \frac{N_{S \nabla 0, 1}}{N_{S0}} \Big] ({\mu_0}_{S\nabla_{\mathcal{B}}} - {\mu_1}_{S\nabla_{\mathcal{B}}}) + \mathbb E \big[ \text{FDP}_0 \big] ({\mu_0}_{S\nabla_{\mathcal{A}}}-{\mu_0}_{S\nabla_{\mathcal{B}}}) - \mathbb E \big[ \text{FDP}_1 \big] ({\mu_1}_{S\nabla_{\mathcal{A}}}-{\mu_1}_{S\nabla_{\mathcal{B}}}).$$ Note that $\widehat{\zeta}_{{N_{S}}}$ only depends on information from $\mathcal{A}$ while $\widehat{\xi}_{{N_{S}}}$ depends on $\mathcal{A}$ and on $\mathcal{B}$. From assumptions \ref{ass linkage random} and \ref{ass delta separated}, $\boldsymbol{X} \indep \Delta$ and $\boldsymbol{X} \indep \Delta |S$ and the potential outcome means are the same among correctly and incorrectly linked profiles (${\mu_t}_{S\nabla} = {\mu_t}_{S\Delta} = {\mu_t}_{\mathcal{S}}$). Besides, $\boldsymbol{X} \not\!\!\!\indep S$ and the impact of linkage errors comes from that dependence. Therefore, $$\mathbb E [ \widehat{\xi}_{{N_{S}}} ] = \underset{\text{opposite contributions}}{\underbrace{\mathbb E \Big[ \frac{N_{S \nabla 1, 0}}{N_{S1}} + \frac{N_{S \nabla 0, 1}}{N_{S0}} \Big] ({\mu_0}_{\mathcal{S}_{\mathcal{B}}} - {\mu_1}_{\mathcal{S}_{\mathcal{B}}})}} + \underset{\text{information discrepancy}}{\underbrace{\mathbb E \big[ \text{FDP}_0 \big] \vphantom{\frac{N_{S \nabla 1, 0}}{N_{S1}}} ({\mu_0}_{\mathcal{S}_{\mathcal{A}}}-{\mu_0}_{\mathcal{S}_{\mathcal{B}}}) - \mathbb E \big[ \text{FDP}_1 \big] ({\mu_1}_{\mathcal{S}_{\mathcal{A}}}-{\mu_1}_{\mathcal{S}_{\mathcal{B}}})}}.$$ 

Moreover, under assumptions \ref{ass linkage random} and \ref{ass most fp}, the proportion of opposite contributions $\mathbb E [ \frac{N_{S \nabla 1, 0}}{N_{S1}} + \frac{N_{S \nabla 0, 1}}{N_{S0}} ]$ is bounded by $\mathbb E [\text{FDP}]$, with the bound saturated in the edge case of \Cref{ass most fp} (when half of the falsely linked data links records from different treatment groups). We obtain that result developing three steps. 

First, \Cref{ass most fp} implies $$\mathbb E \Big[ \frac{N_{S \nabla t, 1-t}}{{N_{S}}_t} \Big] \leq \frac{1}{2} \mathbb E [\text{FDP}_t].$$ 

Second, we notice that $$\text{FDP}=\frac{FP}{FP+TP}=\frac{N_{S \nabla}}{N_{S}}, \text{ hence } \text{FDP}_t = \frac{N_{S \nabla t}}{{N_{S}}_t} = \frac{N_{S \nabla t, t}+N_{S \nabla t, 1-t}}{{N_{S}}_t}.$$ 

Third, for any $t \in \{0,1\}$ \begin{align*}
    \mathbb E [\text{FDP}_t] = \mathbb E \big[ \mathbb E [\text{FDP}_t|\boldsymbol{X}] \big] & = \mathbb E \Big[ \mathbb E \Big[ \frac{\sum_{i,j} S_{i,j} (1-\Delta_{i,j})}{\sum_{i,j} S_{i,j}}|T=t,\boldsymbol{X}\Big] \Big] \\
    & = \mathbb E \Big[ \mathbb E \Big[ \frac{\sum_{i,j} S_{i,j} (1-\Delta_{i,j})}{\sum_{i,j} S_{i,j}}|\boldsymbol{X}\Big] \Big] = \mathbb E \big[ \mathbb E [\text{FDP}|\boldsymbol{X}] \big] = \mathbb E [\text{FDP}]
\end{align*} since $(\boldsymbol{X}, T) \indep \Delta$ by \Cref{ass linkage random} and since $S \indep T | \boldsymbol{X}$. 

As a result, we can write: \begin{align}
    \mathbb E [\widehat{\tau}_{\mathcal{S}}] & = \tau_{\mathcal{A}} + \tau_{\mathcal{S}_{\mathcal{A}}} - \tau_{\mathcal{A}} - \tau_{\mathcal{S}_{\mathcal{B}}} \, \mathbb E \Big[ \frac{N_{S \nabla 1, 0}}{N_{S1}} + \frac{N_{S \nabla 0, 1}}{N_{S0}} \Big] + \mathbb E \big[ \text{FDP}_0 \big] ({\mu_0}_{\mathcal{S}_{\mathcal{A}}}-{\mu_0}_{\mathcal{S}_{\mathcal{B}}}) - \mathbb E \big[ \text{FDP}_1 \big] ({\mu_1}_{\mathcal{S}_{\mathcal{A}}}-{\mu_1}_{\mathcal{S}_{\mathcal{B}}}) \nonumber \\
    & = \tau_{\mathcal{S}_{\mathcal{A}}} - \tau_{\mathcal{S}_{\mathcal{B}}} \, \mathbb E \Big[ \frac{N_{S \nabla 1, 0}}{N_{S1}} + \frac{N_{S \nabla 0, 1}}{N_{S0}} \Big] + \mathbb E [ \text{FDP} ] ({\mu_0}_{\mathcal{S}_{\mathcal{A}}}-{\mu_0}_{\mathcal{S}_{\mathcal{B}}}) - \mathbb E [ \text{FDP} ] ({\mu_1}_{\mathcal{S}_{\mathcal{A}}}-{\mu_1}_{\mathcal{S}_{\mathcal{B}}}) \nonumber\\
    & = \tau_{\mathcal{S}_{\mathcal{A}}} - \tau_{\mathcal{S}_{\mathcal{B}}} \, \mathbb E \Big[ \frac{N_{S \nabla 1, 0}}{N_{S1}} + \frac{N_{S \nabla 0, 1}}{N_{S0}} \Big] + \mathbb E [ \text{FDP} ] (\tau_{\mathcal{S}_{\mathcal{B}}} - \tau_{\mathcal{S}_{\mathcal{A}}}) \nonumber\\
    & = \tau_{\mathcal{S}_{\mathcal{A}}} (1 - \mathbb E [ \text{FDP} ]) + \tau_{\mathcal{S}_{\mathcal{B}}} \underset{\in [0, \mathbb E [ \text{FDP} ]]}{\underbrace{\Big( \mathbb E [ \text{FDP} ] - \mathbb E \Big[ \frac{N_{S \nabla 1, 0}}{N_{S1}} + \frac{N_{S \nabla 0, 1}}{N_{S0}} \Big] \Big)}} \label{final clear result}
\end{align} meaning $\widehat{\tau}_{\mathcal{S}}$ is attenuated compared to $\tau_{\mathcal{S}_{\mathcal{A}}}$ by $\mathbb E [ \text{FDP} ]$, attenuation that is compensated up to $\mathbb E [ \text{FDP} ]$ through the contribution of $\tau_{\mathcal{S}_{\mathcal{B}}}$. 

As noted in \Cref{ass same cov selec} under \Cref{ass linkage random}, distributions of covariates are the same in the selected records from $\mathcal{A}$ and $\mathcal{B}$, thus the information discrepancy disappears on average hence for any $t \in \{0,1\}$, ${\mu_t}_{\mathcal{S}_{\mathcal{A}}}={\mu_t}_{\mathcal{S}_{\mathcal{B}}}$ and $\tau_{\mathcal{S}_{\mathcal{A}}}=\tau_{\mathcal{S}_{\mathcal{B}}}=\tau_{\mathcal{S}}$. Thereby, \Cref{final clear result} becomes $$\mathbb E [\widehat{\tau}_{\mathcal{S}}] = \tau_{\mathcal{S}} \Big( 1 - \mathbb E \Big[ \frac{N_{S \nabla 1, 0}}{N_{S1}} + \frac{N_{S \nabla 0, 1}}{N_{S0}} \Big] \Big),$$ which in the edge case of \Cref{ass most fp} (the worst case) writes: $$\mathbb E [\widehat{\tau}_{\mathcal{S}}] = \tau_{\mathcal{S}} (1 - \mathbb E [ \text{FDP} ]).$$

\subsubsection{The effect on the baseline population is recoverable from sampling bias} \label{sec recoverability appendix}

The causal effect is $S$-recoverable from the record linkage induced selection mechanism $S$ presented in \Cref{fig:SelectionDiagram} under Assumptions \ref{key ass hold} and \ref{ass fdp}. Indeed, point $(i)$ of the selection backdoor criterion \Cref{S-admi} holds under \Cref{key ass hold} discussed in \Cref{sec exchangeability}; point $(ii)$ holds under \Cref{ass fdp} or alternatively under the point of \Cref{rmk link var valid}, we elaborated on that point in \Cref{sec recoverability}; point $(iii)$ and point $(iv)$ hold by definition of the causal record linkage task, as $\boldsymbol{X}$ is measured in both, the selected (biased) population and the baseline population. 

The potential outcome distribution is $S$-recoverable from the selection biased data and the causal effect in the baseline is identified through the following transport formula (see \citep{BareinboimTianPearl2014, pearlbareinboim2014, bareinboimpearl2016}): \begin{align}
    p_\mathcal{A} (Y_i(t)) & = \sum_{\boldsymbol{x}} p_{\mathcal{A}} ( Y_i(t) | \boldsymbol{X}_i=\boldsymbol{x} ) p_{\mathcal{A}} (\boldsymbol{X}_i=\boldsymbol{x}) \label{y distr by def}\\
    & = \sum_{\boldsymbol{x}} p_{\mathcal{S}} ( Y_i(t) | \boldsymbol{X}_i=\boldsymbol{x} ) p_\mathcal{A} (\boldsymbol{X}_i=\boldsymbol{x}), \label{by S admi}
\end{align} which states that we can identify the causal effect in the baseline population through the effect in the specific selected population $\mathcal{S}$.  

By conditional exchangeability and, by distributional consistency (under \Cref{ass fdp} or alternatively under the point of \Cref{rmk link var valid}), the distribution in the selected sample further develops: \begin{align}
    p_\mathcal{A} (Y_i(t)) & = \sum_{\boldsymbol{x}} p_{\mathcal{S}} ( Y_i(t) | \boldsymbol{X}_i=\boldsymbol{x} ) p_\mathcal{A} (\boldsymbol{X}_i=\boldsymbol{x}) \\
    & = \sum_{\boldsymbol{x}} p_{\mathcal{S}} ( Y_i(t) | T_i=t, \boldsymbol{X}_i=\boldsymbol{x} ) p_\mathcal{A} (\boldsymbol{X}_i=\boldsymbol{x}) \label{by cond exch} \\
    & = \sum_{\boldsymbol{x}} p_{\mathcal{S}} ( Y_j | T_i=t, \boldsymbol{X}_i=\boldsymbol{x} ) p_\mathcal{A} (\boldsymbol{X}_i=\boldsymbol{x}). \label{by consistency}
\end{align} From line (\ref{by consistency}) we can estimate the causal effect in the baseline population by estimating the effect in the specific selected population $\mathcal{S}$ and average it over $\boldsymbol{X}$ in the baseline $\mathcal{A}$.

Finally, under \Cref{ass random internal lack pos}, an estimator of $p_{\mathcal{S}} ( Y_j | T_i=t, \boldsymbol{X}_i=\boldsymbol{x} )$ over $\boldsymbol{x} \in \mathcal{S}$ could approximate $p_{\mathcal{A}} ( Y_i(t) | \boldsymbol{X}_i=\boldsymbol{x} )$ over $\boldsymbol{x} \in \mathcal{A}$, even when the covariate distributions in $\mathcal{S}$ and $\mathcal{A}$ differ. Indeed, the lack of overlap can be encompassed under random internal violations of positivity via interpolation---smoothing over the surrounded region of non-positivity---though extrapolation is ill-advised \citep{com_positivity, comp_gmethogs_randpos}. This last assumption guarantees that the resultant expression is estimable from the available data \citep{Correa_Bareinboim_2017}. Hence, $$ \tau_\mathcal{A} = \mathbb E_{\mathcal{A}}[\tau_\mathcal{S}(\boldsymbol{X})] $$ may be recovered using methods for generalised treatment effect estimation.

To recover the effect in practice though, we need to eliminate the biases developed in Appendix \Cref{sec sampling bias} and \Cref{sec attenuation bias} (i.e.\ we need to ensure \Cref{ass fdp}), before transporting the effect estimated in $\mathcal{S}$ to $\mathcal{A}$ as justified here in Appendix \Cref{sec recoverability appendix}. 

\end{appendix}

\section{Author contributions statement}

KR carried out the investigation conceptualised by SP. All authors drew up the simulation framework. KR implemented the estimation methods and conducted the data application and the analysis. She prepared the first manuscript draft, reviewed and edited by MH, MW and SP. All authors read and approved this manuscript.

\section{Acknowledgements}

The authors thank the anonymous reviewers for their valuable suggestions.

\section{Data availability}

The experiments and the data sets are available on \href{https://github.com/robachowyk/CausalRL-experiments}{GitHub}.

\section{Funding}

This research received no specific grant from any funding agency in the public, commercial, or not-for-profit sectors.

\section{Conflict of interest}

None declared.

\bibliographystyle{plainnat}
\bibliography{reference}

\end{document}